\documentclass{article}
\pdfoutput=1

\usepackage{wrapfig}
\usepackage{amsfonts}
\usepackage{amsmath}
\usepackage{amsthm}
\usepackage{amssymb}
\usepackage{graphicx}
\usepackage{textcomp}

\usepackage{tikz}
\usetikzlibrary{arrows}
\usepgflibrary{arrows}

\newcommand{\mathpic}[2]{\ensuremath{\vcenter{\hbox{\begin{tikzpicture}[scale=#1,>=latex'] #2 \end{tikzpicture}}}}}

\newtheorem{theorem}{Theorem}[section]

\newtheorem{remark}[theorem]{Remark}
\newtheorem{example}[theorem]{Example}

\newtheorem{definition}[theorem]{Definition}

\def\be{\begin{eqnarray}}
\def\ee{\end{eqnarray}}

\newcommand{\ZG}{\mathbb{Z}}

\newcommand{\K}{\mathcal{K}}

\newcommand{\br}[1]{\left(#1\right)}

\newcommand{\summ}[2]{\displaystyle \mathop{\sum}_{#1}^{#2}}

\newcommand{\Br}{\mathcal{B}}

\newcommand{\eq}[1]{\begin{equation} #1 \end{equation}}
\newcommand{\ex}[1]{\begin{example} #1 \end{example}}
\newcommand{\dfn}[1]{\begin{definition} #1 \end{definition}}

\newcommand{\eqlm}[2]{\begin{multline} \label{e:#1} #2 \end{multline}}

\newcommand{\eql}[2]{\begin{equation} \label{e:#1} #2 \end{equation}}
\newcommand{\dfnl}[2]{\begin{definition} \label{e:#1} #2 \end{definition}}

\newcommand{\eqs}[1]{\begin{align} #1 \end{align}}

\newcommand{\re}[1]{(\ref{e:#1})}

\newcommand{\ths}{\vartheta}

\newcommand{\ep}{\varepsilon}

\newcommand{\ts}[1]{\ifcase#1 \ths_{\br{D_8\oplus D_8}^+}\or \ths_{\ZG\oplus A_{15}^+}\or\ths_{\ZG^2\oplus \br{E_7\oplus E_7}^+}\or\ths_{\ZG^4\oplus D_{12}^+}\or \ths_{\ZG^8\oplus E_{8}}\or \ths_{\ZG^{16}} \or \ths_{E_8 \oplus E_8} \or \ths_{D_{16}^+} \fi}
\DeclareMathOperator{\tr}{tr}

\newcommand{\ffac}{\mathpic{0.6}{\draw[gray] (0,0) circle (1);
\draw[red,very thick] (0,1) -- (0,-1);
\draw[red,very thick] (1,0) -- (-1,0);
}}
\newcommand{\ffab}{\mathpic{0.6}{\draw[gray] (0,0) circle (1);
\draw[red,very thick] (0,1) -- (1,0);
\draw[red,very thick] (0,-1) -- (-1,0);
}}
\newcommand{\ffaa}{\mathpic{0.6}{\draw[gray] (0,0) circle (1);
\draw[red,very thick] (0,1) -- (0,-1);
}}

\newcommand{\oau}{\mathpic{0.6}{\draw[gray] (0,0) circle (1); \draw[->,red,very thick] (0,1) -- (0.,-1.);}}

\newcommand{\uao}{\mathpic{0.6}{\draw[gray] (0,0) circle (1); \draw[->,red,very thick] (0,-1) -- (0.,1.);}}

\newcommand{\ouuo}{\mathpic{0.6}{\draw[gray] (0,0) circle (1); \draw[->,red,very thick] (0,1) -- (0.,-1.); \draw[->,red,very thick] (1,0) -- (-1.,0.);}}

\newcommand{\uoou}{\mathpic{0.6}{\draw[gray] (0,0) circle (1); \draw[->,red,very thick] (0,-1) -- (0.,1.); \draw[->,red,very thick] (-1,0) -- (1.,0.);}}

\newcommand{\tappp}{\mathpic{0.6}{\draw[gray] (0,0) circle (1); \draw[->,blue,very thick,dashdotted] (0,1) -- (0.,-1.); \draw[->,blue,very thick,dashdotted] (0.866025,0.5) -- (-0.866025,0.5); \draw[->,blue,very thick,dashdotted] (0.866025,-0.5) -- (-0.866025,-0.5);}}

\newcommand{\chdiag}{\mathpic{0.4}{\draw[black] (0,0) circle (1); \draw[black] (0,1) -- (0.,-1.); \draw[black] (0.866025,0.5) -- (-0.866025,0.5); \draw[black] (0.866025,-0.5) -- (-0.866025,-0.5);}}

\newcommand{\tappm}{\mathpic{0.6}{\draw[gray] (0,0) circle (1); \draw[->,blue,very thick,dashdotted] (0,1) -- (0.,-1.); \draw[->,blue,very thick,dashdotted] (0.866025,0.5) -- (-0.866025,0.5); \draw[->,blue,very thick,dashdotted] (-0.866025,-0.5) -- (0.866025,-0.5);}}

\newcommand{\tapmp}{\mathpic{0.6}{\draw[gray] (0,0) circle (1); \draw[->,blue,very thick,dashdotted] (0,1) -- (0.,-1.); \draw[->,blue,very thick,dashdotted] (-0.866025,0.5) -- (0.866025,0.5); \draw[->,blue,very thick,dashdotted] (0.866025,-0.5) -- (-0.866025,-0.5);}}

\newcommand{\tbpmp}{\mathpic{0.6}{\draw[gray] (0,0) circle (1);
\draw[->,blue,very thick,dashdotted] (0,1) -- (-0.866025,-0.5);
\draw[<-,blue,very thick,dashdotted] (0.866025,0.5) -- (0,-1);
\draw[->,blue,very thick,dashdotted] (0.866025,-0.5) -- (-0.866025,0.5);
}}

\newcommand{\tbpmm}{\mathpic{0.6}{\draw[gray] (0,0) circle (1);
\draw[->,blue,very thick,dashdotted] (0,1) -- (-0.866025,-0.5);
\draw[<-,blue,very thick,dashdotted] (0.866025,0.5) -- (0,-1);
\draw[<-,blue,very thick,dashdotted] (0.866025,-0.5) -- (-0.866025,0.5);
}}

\newcommand{\tcppp}{\mathpic{0.6}{\draw[gray] (0,0) circle (1);
\draw[->,blue,very thick,dashdotted] (0,1) -- (0.866025,-0.5);
\draw[->,blue,very thick,dashdotted] (0.866025,0.5) -- (-0.866025,-0.5);
\draw[->,blue,very thick,dashdotted] (0,-1) -- (-0.866025,0.5);
}}

\newcommand{\tcpmp}{\mathpic{0.6}{\draw[gray] (0,0) circle (1);
\draw[->,blue,very thick,dashdotted] (0,1) -- (0.866025,-0.5);
\draw[<-,blue,very thick,dashdotted] (0.866025,0.5) -- (-0.866025,-0.5);
\draw[->,blue,very thick,dashdotted] (0,-1) -- (-0.866025,0.5);
}}

\newcommand{\tcppm}{\mathpic{0.6}{\draw[gray] (0,0) circle (1);
\draw[->,blue,very thick,dashdotted] (0,1) -- (0.866025,-0.5);
\draw[->,blue,very thick,dashdotted] (0.866025,0.5) -- (-0.866025,-0.5);
\draw[<-,blue,very thick,dashdotted] (0,-1) -- (-0.866025,0.5);
}}

\newcommand{\tdpmp}{\mathpic{0.6}{\draw[gray] (0,0) circle (1);
\draw[->,blue,very thick,dashdotted] (0,1) -- (0,-1);
\draw[<-,blue,very thick,dashdotted] (0.866025,0.5) -- (-0.866025,-0.5);
\draw[->,blue,very thick,dashdotted] (0.866025,-0.5) -- (-0.866025,0.5);
}}

\newcommand{\oap}{\mathpic{0.6}{\draw[gray] (0,0) circle (1);
\draw[->,blue,very thick,dashdotted] (0,1) -- (0,-1);
}}

\newcommand{\tapp}{\mathpic{0.6}{\draw[gray] (0,0) circle (1);
\draw[->,blue,very thick,dashdotted] (0,1) -- (0,-1);
\draw[->,blue,very thick,dashdotted] (1,0) -- (-1,0);
}}

\newcommand{\tapm}{\mathpic{0.6}{\draw[gray] (0,0) circle (1);
\draw[->,blue,very thick,dashdotted] (0,1) -- (0,-1);
\draw[<-,blue,very thick,dashdotted] (1,0) -- (-1,0);
}}

\newcommand{\tbpp}{\mathpic{0.6}{\draw[gray] (0,0) circle (1);
\draw[->,blue,very thick,dashdotted] (0,1) -- (-1,0);
\draw[->,blue,very thick,dashdotted] (1,0) -- (0,-1);
}}

\newcommand{\tbpm}{\mathpic{0.6}{\draw[gray] (0,0) circle (1);
\draw[->,blue,very thick,dashdotted] (0,1) -- (-1,0);
\draw[<-,blue,very thick,dashdotted] (1,0) -- (0,-1);
}}

\newcommand{\tcpp}{\mathpic{0.6}{\draw[gray] (0,0) circle (1);
\draw[->,blue,very thick,dashdotted] (0,1) -- (1,0);
\draw[->,blue,very thick,dashdotted] (0,-1) -- (-1,0);
}}

\newcommand{\tcpm}{\mathpic{0.6}{\draw[gray] (0,0) circle (1);
\draw[->,blue,very thick,dashdotted] (0,1) -- (1,0);
\draw[->,blue,very thick,dashdotted] (0,-1) -- (-1,0);
}}

\hoffset= - 1.3in         

\title{{\bf Kontsevich integral for knots and 
Vassiliev invariants} \vspace{.2cm}}
\author{{\bf P.Dunin-Barkowski}\thanks{{\small
{\it ITEP, Moscow, Russia and Korteweg-de Vries Institute for Mathematics, University of Amsterdam, The Netherlands}};
barkovs@itep.ru}, {\bf A.Sleptsov}\thanks{{\small
{\it ITEP, Moscow, Russia}};
sleptsov@itep.ru}, {\bf A.Smirnov}\thanks{{\small
{\it ITEP, Moscow, Russia and Columbia University, New York, USA}}; asmirnov@itep.ru}
\date{ }}

\begin{document}
\maketitle
\vspace{-5.0cm}
\begin{center}
\hfill ITEP/TH-63/11\\
\end{center}

\vspace{3.5cm}

\bigskip

\centerline{ABSTRACT}

\bigskip

{\footnotesize
We review quantum field theory approach to the knot theory. Using holomorphic gauge we obtain the Kontsevich integral. It is explained how to calculate Vassiliev invariants and coefficients in Kontsevich integral in a combinatorial way which can be programmed on a computer. We discuss experimental results and temporal gauge considerations which lead to representation of Vassiliev invariants in terms of arrow diagrams. Explicit examples and computational results are presented.
}

\bigskip

\tableofcontents

\newpage
\section{Introduction}
It is believed that the path-integral representation for knot invariants arising
from topological quantum field theory (QFT) gives the most profound and general
description of knot invariants. Ideologically, it means that all possible descriptions of
knot invariants can be derived from this representation by utilizing different methods
of path-integral calculus. For example, the usage of certain non-perturbative
methods leads to well known description of polynomial knot invariants through the "skein relations"
\cite{WCS}. The perturbative computations naturally lead to the numerical
Vassiliev Invariants \cite{Bar3}. In the last case we obtain the formulae for Vassiliev invariants in the form of ``Feynman integrals''. For a recent comprehensive treatise on Vassiliev invariants see \cite{ChDuBook}.

Despite beauty and simplicity of this picture many problems in the theory of knot invariants remain unsolved. Currently more mathematical descriptions of knot invariants are developed than can be derived from path-integral. One such problem is the derivation of quantum group invariants from the path-integral representation.
The main ingredient in the theory of these invariants is the universal quantum $R$-matrix defined for integrable quantum deformation of a Lie group. The appearance of the quantum groups in the path-integral representation looks mysterious and we lack the derivation of the corresponding $R$-matrix (object with noncommutative matrix elements should appear from classical integral) from path-integral. These problems were discussed in details in \cite{Sm1,Sm2}.

The second interesting problem (which as we believe is closely connected to the first one) concerns the combinatorial description of the numerical Vassiliev invariants of knots. At the moment, there exist three different description of Vassiliev invariants: through the generalized Gauss
integrals \cite{Gu10}, through Kontsevich integral \cite{Konts1}, and finally there are combinatorial
formulae for Vassiliev invariants of orders 2,3 and 4, \cite{PV1}-\cite{La5}. The first two descriptions can be easily derived from
the path-integral but (surprisingly) we lack such a derivation for combinatorial formulae.
The Gauss integral representation for Vassiliev invariants comes from the perturbative computations
of path-integral in the covariant Lorentz gauge. Similarly, usage of non-covariant holomorphic
gauge (which is sometimes referred to as light-cone gauge) computation of path-integral leads to Kontsevich integral \cite{La6}.
The aim of this paper is to show that the universal combinatorial formulae for Vassiliev invariants
and quantum group invariants can be derived from path-integral in the temporal gauge.

The main difference between the non-covariant gauges (temporal or holomorphic gauge) and covariant Lorentz gauge is that the Feynman integrals in there gauges can be naturally "localized", in the sense that only a finite number of special points on the knots contribute to the integrals. In the Lorentz gauge the Feynman integrals have a form of multiple 3-d integrals (see (\ref{gi1}),(\ref{gi2}) as an example) and all points of the knot enter the integral equally. On the contrary, in the holomorphic gauge only special points contribute to the Feynman integrals. Summed in all orders of perturbation theory these contributions lead to definition of polynomial knot invariants through simple crossing operator and Drinfeld associator with rational zeta-function coefficients \cite{ChDu}. We discuss this localization process in detail in Section \ref{KIP}.

Before we pass to temporal gauge consideration let us briefly describe Section \ref{numvas}, where the computation of Vassiliev invariants is discussed. These results are used in Section \ref{TG} for numerical studies. To compute Vassiliev invariants we use technique developed in previous Section \ref{KIP}. With the help of it we can calculate coefficients in Kontsevich integral expansion in chord diagrams. However, these coefficients are not invariants due to chord diagrams being not independent. We choose a basis among the independent ones with the help of so-called trivalent diagrams. Thus we can find Vassiliev invariants. In the rest of the Section we describe two polynomials in Vassiliev invariants up to level 6 vanishing on a large class of considerable knots.

The temporal gauge is distinguished among all gauges. The Feynman integrals here have ultra-local form -- only crossing points of two-dimensional projection of the knot contribute to the answer.
This is exactly what happens in the quantum group description of knot invariants where the crossing points contribute as a universal quantum $R$-matrix.
On the other hand, the combinatorial formulae for Vassiliev invariants also are based on the information from these crossing points only.
All these observations make it natural to argue that the Feynman integrals arising from the
general path-integral representation in the temporal gauge should give the universal combinatorial
 formulae for Vassiliev invariants, and summed in all orders of perturbation theory they should give the perturbative
  expansion of the universal quantum $R$-matrix for $\hbar$-deformed gauge group of the path integral.
  These questions and structures arising from path-integral representation of knot invariants in the temporal gauge
   are discussed in Section \ref{TG}.

As a side remark, recently new polynomial invariants of knots appeared, the so-called ``superpolynomials'' \cite{Gukov,DBMMSS}, which are generalizations of HOMFLY polynomials. However, there is still no related QFT-like formulation as the one discussed in the present paper. Probably this interpritation should involve something like ``Chern-Simons theory with a quantum group'' which is still not constructed (for an attempt see \cite{AgSh}).

\section{Knot invariants from Chern-Simons theory}
\label{kics}

The incorporation of Vassiliev invariants in the path-integral representation is clear from the following picture.
Let $A$ be a connection on $\mathbb{R}^3$ taking values in some
representation $R$ of a Lie algebra $g$, i.e., in components:
$$
A=A_{i}^{a}(x)\,T^{a}\,dx^{i},
$$
where $T^{a}$ are the generators of $g$.
Let curve $C$ in $\mathbb{R}^3$ give a particular realization of knot $K$.
Consider the holonomy of $A$ along $C$, it is given by
the ordered exponent:
$$
\Gamma(C,A) = P \exp \oint\limits_{C} A = 1+ \oint\limits_{C} A_{i}^{a}(x) T^{a}+
\oint\limits_{C} A_{i_{1}}^{a_{1}}(x_{1})\int\limits_{0}^{x_{1}} A_{i_{2}}^{a_{2}}(x_{2})
\, T^{a_{1}} T^{a_{2}}+...
$$
The Wilson loop along $C$ is a function depending on $C$ and $A$ defined as a trace of holonomy:
$$ W_R(C,A)=\tr_R \Gamma(C,A) $$
According to \cite{WCS} there exists a functional $S_{CS} (A)$
(we write it down explicitly later) such that the integral averaging of the Wilson loop with the
weight $\exp\Big(-\frac{2\pi i}{\hbar} S(A)\Big)$  has the following remarkable property:
\be
\label{W}
\begin{array}{|c|}
\hline\\
\ \ \  \langle \, W_R(K)\,\rangle= \dfrac{1}{Z} \displaystyle \int DA \, \exp\Big(-\frac{2\pi i}{\hbar} S_{CS}(A)\Big)\, W_R(C,A) \ \ \  \\
\\
\hline
\end{array}
\ee
where
$$
 Z=\int DA\,\exp\Big(-\frac{2\pi i}{\hbar} S_{CS}(A) \Big)
$$
i.e. the averaging of $W_{R}(C,A)$ with the weight $\exp\Big(-\frac{2\pi i}{\hbar} S(A)\Big)$
does not depend on the realization $C$ of the knot in $\mathbb{R}^3$
but only on the topological class of equivalence of knot $K$ (in what follows we will denote
the averaging of quantity $Q$ with this weight by $\langle Q\rangle$) and therefore,  $\langle\, W(K)\,\rangle$ defines a knot invariant.

The distinguished Chern-Simons action giving the invariant average (\ref{W}) has the following form:
\be
\label{CSA}
S_{CS}(A)=\int \limits_{\mathbb{R}^3}\,\tr ( A\wedge dA +\dfrac{2}{3} A\wedge A\wedge A )
\ee
If we normalize the algebra generators $T^{a}$ as $\tr (T^{a}T^{b})=\delta^{a b}$ and define the structure constants $f$ of algebra $g$ as $[T^{a},T^{b}]=f_{a b c}\,T^{c}$  then the action takes the form:
$$
S_{CS}(A)=\epsilon^{i j k} \int\limits_{\mathbb{R}^3} dx^{3}\, A^{a}_{i} \partial_{j} A_{k}^{a} +\dfrac{1}{6} f_{a b c} A^{a}_{i}A^{b}_{j}A^{c}_{k}
$$

Formula (\ref{W}) is precisely the path integral representation of knot invariants. It is believed that all invariants of knots can be derived from this expression. Let us outline the appearance of Vassiliev invariants in this scheme. Obviously the mean value $\langle\,W(K)\,\rangle$ has the following structure:
\be
\label{eq1}
\nonumber
 \langle\, W(C,A)\, \rangle = \langle \,\sum\limits_{n=0}^{\infty} \oint dx_{1}\int
dx_{2}...\int dx_{n} A^{a_1}(x_{1})A^{a_2}(x_{2})...A^{a_3}(x_{n})\,
\tr(T^{a_1} T^{a_2}...T^{a_n}) \, \rangle=\\
=\sum\limits_{n=0}^{\infty} \oint dx_{1}\int
dx_{2}...\int dx_{n} \langle\,A^{a_1}(x_{1})A^{a_2}(x_{2})...A^{a_3}(x_{n})\,\rangle\,
\tr(T^{a_1} T^{a_2}...T^{a_n}) =
\sum\limits_{n=0}^{\infty} \sum \limits_{m=1}^{N_{n}} V_{n,m}\, G_{n,m}
 \ee
From this expansion we see that the information about the knot and the gauge
group enter $<W(K)>$ separately. The information about the embedding of knot into
$\mathbb{R}^3$ is encoded in the integrals of the form:
$$
V_{n,m} \sim \oint dx_{1}\int
dx_{2}...\int dx_{n} \langle\,A^{a_1}(x_{1})A^{a_2}(x_{2})...A^{a_3}(x_{n})\,\rangle
$$
and the information about the gauge group and representation enter the answer as the "group factors":
$$
G_{n,m} \sim \tr(T^{a_1} T^{a_2}...T^{a_n})
$$
$G_{k,m}$ are the group factors called \textit{chord diagrams with $n$ chords}. Chord diagrams with $n$ chords form a vector space of dimension $N_{n}$.
Despite $\langle W(K) \rangle$ being a knot invariant, the numbers $V_{n,m}$ are not invariants. This is because
the group elements $G_{n,m}$ are not independent, and the coefficients $V_{n,m}$ are invariants only up to relations among $G_{n,m}$. In order to pass to the the knot
invariants we need to choose in the space of group elements some basis of independent
group elements and expand the sum (\ref{eq1}) in this basis. Finally as it was proven in
\cite{La8} we arrive to the following infinite product formula:
\be
\label{eq2}
\begin{array}{|c|}
\hline \\
\ \ \  \langle\,W(K)\,\rangle=d_{R}\,\prod\limits_{n=1}^{\infty} \prod\limits_{m=1}^{{\cal{N}}_{n}} \exp( \hbar^{n} {\cal{V}}_{n,m} {\cal{G}}_{n,m}) \ \ \ \\
\\
\hline
\end{array}
\ee
where ${\cal{N}}_{n}$ is the number of independent group elements of degree $n$ and ${\cal{V}}_{n,m}$
are the coefficients of independent group factors ${\cal{G}}_{n,m}$ which are nothing
but the Vassiliev invariants of degree $n$; $d_{R}$ stands for the dimension of representation $R$. Note that the number of independent
Vassiliev invariants of degree $n$ is given by the number of independent group
elements ${\cal{N}}_{n}$. We list the first several values of ${\cal{N}}_{n}$ :
\begin{equation}
\label{tab}
\begin{array}{|c|c|c|c|c|c|c|}
\hline
n&1&2&3&4&5&6\\
\hline
{\cal{N}}_{n}&1&1&1&2&3&5\\
\hline
\end{array}
\end{equation}
Formula (\ref{eq2}) along with table (\ref{tab}) means that the expansion of
$\langle\,W(K)\,\rangle$ up to order $3$ is the following one:
$$
\frac{1}{d_R} \langle\,W(K)\,\rangle= 1 + \hbar {\cal{V}}_{1,1} {\cal{G}}_{1,1}  +
\hbar^2 \Big( \frac{1}{2!} {\cal{V}}_{1,1}^{2} {\cal{G}}^{2}_{1,1} +{\cal{V}}_{2,1} {\cal{G}}_{2,1}   \Big)+
\hbar^3\Big( \frac{1}{3!} {\cal{V}}_{1,1}^{3} {\cal{G}}^{3}_{1,1}
+{\cal{V}}_{1,1} {\cal{V}}_{2,1} {\cal{G}}_{1,1} {\cal{G}}_{2,1}+ {\cal{V}}_{3,1} {\cal{G}}_{3,1} \Big)+...
$$
Note, that here the relations between group factors are taken into account.

QFT provides several techniques for computing coefficients of $\hbar$-expansion for $<W(K)>$, each technique leads to some formulae for Vassiliev invariants. The straightforward way is to use the perturbation theory for covariant Lorentz gauge
$\partial_{i} A_{i}=0$. Standard quantization technique for Lorentz gauge considered in
\cite{Gu10} leads to the following Feynman integral formulae for the first two Vassiliev invariants (integrals are taken along the curve representing the knot):
\be \label{gi1}{\cal{V}}_{1,1}=\oint\limits_{C} dx_{1}^{i} \int_{0}^{x_{1}} dx_{2}^{j} \epsilon_{i j k}
\frac{(x_{1}-x_{2})^{k}}{|x_{1}-x_{2}|^3}  \ee
\be\nonumber
{\cal{V}}_{2,1}=\frac{1}{2} \oint\limits_{C} dx_{1}^{i} \int\limits_{0}^{x_{1}} dx_{2}^{j}
\int\limits_{0}^{x_{2}} dx_{3}^{k} \int\limits_{0}^{x_{3}} dx_{4}^{m}
\epsilon_{p j q} \epsilon_{k i s} \frac{(x_{4}-x_{2})^{q}}{|x_{4}-x_{2}|^3}
\frac{(x_{3}-x_{1})^{s}}{|x_{3}-x_{1}|^3}+ \\
\label{gi2}
-\frac{1}{8} \oint\limits_{C} dx_{1}^{i} \int\limits_{0}^{x_{1}} dx_{2}^{j}
\int\limits_{0}^{x_{2}} dx_{3}^{k} \int\limits_{\mathbb{R}^3} dx_{4}^{3}
\epsilon^{p r s} \epsilon_{i p m} \epsilon_{j r n} \epsilon_{k s t}
\dfrac{(x_{4}-x_{1})^m}{|x_{4}-x_{1}|^3} \dfrac{(x_{4}-x_{2})^n}{|x_{4}-x_{2}|^3}
\dfrac{(x_{4}-x_{3})^t}{|x_{4}-x_{3}|^3}
\ee
The first integral here is the so called Gauss integral for self linking number.
In general, perturbation theory in Lorentz gauge provides
the theory of Vassiliev invariants in terms of rather sophisticated
$3-d$ "generalized Gauss integrals". Therefore we will consider different choices of gauge in the following Sections.

\section{Holomorphic gauge}
\label{KIP}
The aim of this Section is to show that the universal Kontsevich integral for knots can be obtained from the general path integral representation. In fact, as it was explained in \cite{La6} the Kontsevich integral appears as average value $\langle\,W(K)\,\rangle $ computed in the so called holomorphic gauge. To show it, let us start with decomposition of three-dimensional space $\mathbb{R}^3=\mathbb{R}\times \mathbb{C}$, i.e. we pass from the coordinates $(x_{0},x_{1},x_{2})$ to $(t,z,\bar{z})$ defined as:
$$
t=x_{0},\ \ \ z=x_{1}+i x_{2},\ \ \ \bar{z}=x_{1}-i x_{2}
$$
then the for differentials and dual bases we have:
\be
\nonumber
d z=d x_{1}+i d x_{2},\ \ \ \partial_{z}=\frac{1}{2}(\partial_{x_{1}}-i \partial_{x_{2}})\\
\nonumber
d \bar{z}=d x_{1}-i d x_{2},\ \ \ \partial_{\bar{z}}=\frac{1}{2}(\partial_{x_{1}}+i \partial_{x_{2}})
\ee
The gauge field takes the form:
 $$ A^{a}_{i} dx^{i}= A^{a}_{0} dx^{0} + \dfrac{1}{2} A^{a}_{z} dz + \dfrac{1}{2} A^{a}_{\bar{z}} d\bar{z}
 $$
where:
$$
A_{z}^{a}=A_{1}^{a}-i A_{2}^{a},\ \ \
A_{\bar{z}}^{a}=A_{1}^{a}+i A_{2}^{a}
$$
As usual, the path integral (\ref{W}) is not well defined due to gauge symmetry, to make it convergent we need to fix an appropriate gauge condition. The holomorphic gauge is defined by the following non-covariant condition
\be
\label{lcg}
\begin{array}{|c|}
\hline
 A_{\bar{z}}=0\\
 \hline
 \end{array}
\ee
The main feature of this gauge is that the initially complex, cubical Chern-Simons action becomes pure quadratic and the resulting path integral becomes Gaussian. This allows one to compute it perturbatively utilizing the Wick theorem. Indeed, in the gauge (\ref{lcg}) the cubical part of the action (\ref{CSA}) vanishes:
$$
\left.A\wedge A\wedge A\right|_{A_{\bar{z}}=0}=0
$$
and we end up with the following quadratic action\footnote{Of course, the careful gauge fixing involves the Faddeev-Popov procedure which leads to an additional ghost term in the action. Fortunately, in the holomorphic gauge the ghost fields are not coupled to the gauge ones and therefore can be simply integrated away from the path integral. In this way we again arrive to the quadratic action  (\ref{CSAlcg}).}:
\be
\label{CSAlcg}
\begin{array}{|c|}
\hline\\
\ \ \ \left.S(A)\right|_{A_{\bar{z}}=0}=i\,\displaystyle\int\limits_{\mathbb{R}^3} dt d\bar{z} dz\,\epsilon^{m n} A_{m}^{a} \partial_{\bar{z}} A_{n}^{a}\ \ \ \\
\\
\hline
\end{array}
\ee
where $\epsilon^{m n}$ is antisymmetric, $m,n \in \{t, z\}$ and $\epsilon^{t z}=1$.
The main ingredient of perturbation theory with action (\ref{CSAlcg}) is the gauge propagator defined as the inverse of quadratic operator of the action:
\be
\label{lcp}
\langle\, A_{m}^{a}(t_{1},z_{1},\bar{z}_{1})\, A_{n}^{b}(t_{2},z_{2},\bar{z}_{2}) \,\rangle =\Big(\frac{\delta^{ab}}{\hbar}\,\epsilon^{n m} \partial_{\bar{z}}\Big)^{-1}= \epsilon_{m n} \,\delta^{a b}\, \dfrac{\hbar}{2\pi i}\dfrac{\delta(t_{1}-t_{2})}{z_{1}-z_{2}}
\ee
To find it we need the following simple fact about the operator $\partial_{\bar{z}}$:
$$
\partial_{\bar{z}}^{-1}=\dfrac{1}{2\pi i} \dfrac{1}{z}
$$
To prove it, we note that the inverse of Laplace operator $\Delta = \partial_{z}\partial_{\bar{z}} $ (its Green function) on the complex plane is given by the logarithm function\footnote{Note that $\partial_{\bar{z}}^{-1}$ is defined up to any holomorphic function, as they are in the kernel of $\partial_{\bar{z}}$:
$$
\partial_{\bar{z}}^{-1}=\dfrac{1}{2\pi i} \dfrac{1}{z}+f(z)
$$
The trick with the Laplacian consists of the following: we restrict the operators to the space of functions with absolute values sufficiently fast decreasing at the infinity. The only holomorpfic function with this property is $f(z)=0$.
}
$$
(\partial_{z}\partial_{\bar{z}})^{-1}=\dfrac{1}{2\pi i}\log ( z \bar{z}), \ \ {\rm{therefore}}\ \ \ \partial_{\bar{z}}^{-1}=\dfrac{1}{2\pi i} \partial_{z}\log ( z \bar{z})=\dfrac{1}{2\pi i} \dfrac{1}{z}
$$
Let us show that average value $ \langle W(K) \rangle $ in the holomorphic gauge coincides with the Kontsevich integral for the knot. To proceed we introduce an orientation on a knot in $\mathbb{R}^3$. Obviously, if we pick some point $p$ on the knot then $K\setminus p$ is topologically a segment $I_{p}=(0,1)$. Orientation on $I_{p}$ naturally defines orientation on $K\setminus p$. We will denote this orientation using symbol $o$, for example we can compare two points $x_{1}$ and $x_{2}$ as $o(x_{1})<o(x_{2})$. The Wilson loop is given by ordered exponent which has the following form:
$$
W(K)= \tr P \exp\Big( \oint A_{\mu}^{a}(x) T^{a} dx^{\mu}\Big) = \sum\limits_{n=0}^{\infty} \,\int\limits_{o(x_{1})<o(x_{2})<...o(x_{n})} \prod\limits_{k=1}^{n}\,dx^{\mu_{k}}_{k} \, A_{\mu_{1}}^{a_{1}}(x_{1}) A_{\mu_2}^{a_{2}}(x_{2})...A_{\mu_n}^{a_{n}}(x_{n}) \tr\Big(T^{a_{1}}T^{a_{2}}...T^{a_{n}}\Big)
$$
For the average $\langle\, W(K)\,\rangle$ we get:
$$
\langle\, W(K)\,\rangle=\sum\limits_{n=0}^{\infty} \,\int\limits_{o(x_{1})<o(x_{2})<...o(x_{n})} \prod\limits_{k=1}^{n}\,dx^{\mu_{k}}_{k} \, \langle\, A_{\mu_{1}}^{a_{1}}(x_{1}) A_{\mu_2}^{a_{2}}(x_{2})...A_{\mu_n}^{a_{n}}(x_{n})\,\rangle \tr\Big(T^{a_{1}}T^{a_{2}}...T^{a_{n}}\Big)
$$
 As the action of the theory is quadratic, the average of $n$ fields is not equal to zero only for even $n$, moreover the Wick theorem gives:
\be
\label{wt}
\langle\, A_{\mu_{1}}^{a_{1}}(x_{1}) A_{\mu_2}^{a_{2}}(x_{2})...A_{\mu_{2n}}^{a_{2n}}(x_{2 n})\,\rangle=\sum\limits_{\ \ \ \Big((i_1,j_1) (i_2,j_2)...(i_n,j_n)\Big) \in P_{2n} }\, \prod\limits_{k=1}^{n}\, \langle\, A_{\mu_{i_{k}}}^{a_{i_k}}(x_{i_k})A^{a_{j_k}}_{\mu_{j_{k}}}(x_{i_k})  \,\rangle
\ee
Here the sum runs over the set of all pairings $P_{2n}$ of $2n$ numbers. An element of this set has the form $p=\Big((i_1,j_1) (i_2,j_2)...(i_n,j_n)\Big)$ where $i_{k}<j_{k}$ and the numbers $i_{k}, j_{k}$ are all different numbers from the set $\{1,2,...,2n\}$. If $p\in P_{2n}$ is a pairing then we define a function  $\sigma_{p}$ on a set $\{1,2,...,2n\}$ as follows:
$$
\sigma_{p}(i_{k})=i_{k}, \ \ \ \sigma_{p}(j_{k})=i_{k}
$$
i.e. it returns a minimum number from the pair $(i_{k},j_{k})$. The Wick theorem (\ref{wt}) gives:
$$
\langle\, W(K)\,\rangle=\sum\limits_{n=0}^{\infty} \,\int\limits_{o(x_{1})<o(x_{2})<...o(x_{n})}\,\sum\limits_{p\,\in P_{2n}}\, \prod\limits_{k=1}^{n} dx^{\mu_{i_k}}_{i_k}\,dx^{\mu_{j_k}}_{j_k}\, \langle\, A_{\mu_{i_{k}}}^{a_{i_k}}(x_{i_k})A^{a_{j_k}}_{\mu_{j_{k}}}(x_{i_k})  \,\rangle \, \tr\Big(T^{a_{\sigma_{p}(1)}} T^{a_{\sigma_{p}(2)}}...T^{a_{\sigma_{p}(2n)}}\Big)
$$
From the propagator (\ref{lcp}) we have:
$$
dx^{\mu_{i_k}}_{i_k}\,dx^{\mu_{j_k}}_{j_k}\, \langle\, A_{\mu_{i_{k}}}^{a_{i_k}}(x_{i_k})A^{a_{j_k}}_{\mu_{j_{k}}}(x_{i_k})  \,\rangle = \dfrac{\hbar}{2\pi i} \, \delta^{a_{i_k} a_{j_k}}\,(dz_{i_k} dt_{j_k}-dz_{j_k} dt_{i_k}) \, \dfrac{\delta(t_{i_k}-t_{j_k})}{z_{i_k}-z_{j_k}}
$$
therefore,
 $$
 \langle\, W(K)\,\rangle=\sum\limits_{n=0}^{\infty} \,\dfrac{\hbar^n}{(2\pi i)^n}\,\int\limits_{o(z_{1})<o(z_{2})<...o(z_{n})}\,\sum\limits_{p\,\in P_{2n}}\,\prod\limits_{k=1}^{n} \left((dz_{i_k} dt_{j_k}-dz_{j_k} dt_{i_k}) \, \dfrac{\delta(t_{i_k}-t_{j_k})}{z_{i_k}-z_{j_k}}\right)\, G_{p}
 $$
 where the pairing $p=\Big((i_1,j_1) (i_2,j_2)...(i_n,j_n)\Big)$ and the group element $G_{p}=\tr\Big( T^{a_{\sigma_{p}(1)}} T^{a_{\sigma_{p}(2)}}...T^{a_{\sigma_{p}(2n)}} \Big)$.
To obtain the Kontsevich integral we need to make the last step: integrate out the delta functions $\delta(t_{i_k}-t_{j_k})$ from the integral. To do it, it is convenient to parametrize the knot by the "height" $t$. 
Such parametrization of a knot is regular at non-critical points with respect to $t$ direction. In the interval between the critical points we have a well defined function $z(t)$ which parametrizes the knot in $\mathbb{R}\times \mathbb{C}$. Using this parametrization we rewrite the integral in the form:
 $$
 \langle\, W(K)\,\rangle=\sum\limits_{n=0}^{\infty} \,\dfrac{\hbar^n}{(2\pi i)^n}\,\int\limits_{o(z_{1})<o(z_{2})<...o(z_{n})}\,\sum\limits_{p\,\in P_{2n}}\,(-1)^{p_\downarrow}\,\prod\limits_{k=1}^{n}\,dt_{i_k} dt_{j_k}\,\sum\limits_{p\,\in P_{2n}}\,\left(\dfrac{dz_{i_k}(t_{i_k})}{dt_{i_k} }-\dfrac{dz_{j_k}(t_{j_k})}{dt_{j_k}}\right) \, \dfrac{\delta(t_{i_k}-t_{j_k})}{z_{i_k}-z_{j_k}}\, G_{p}
 $$
 Here we should be careful not to forget the sign factor $(-1)^{p_\downarrow}$ where
 $p_\downarrow$ is the number of down-oriented segments between critical points on the knot entering the integral. On these segments the "height" parameter and the orientation of the knot are opposite, and therefore we should change $dt_{i}\rightarrow - dt_{i} $ which finally results in factor $(-1)^{p_\downarrow}$.
 Integrating the $t_{j_{k}}$ variables we obtain:
 $$
 \langle\, W(K)\,\rangle=\sum\limits_{n=0}^{\infty} \,\dfrac{\hbar^n}{(2\pi i)^n}\,\int\limits_{o(z_{1})<o(z_{2})<...o(z_{n})}\,\sum\limits_{p\,\in P_{2n}}\,(-1)^{p_\downarrow}\,\prod\limits_{k=1}^{n}\,dt_{i_k}\, \left(\dfrac{dz_{i_k}(t_{i_k})}{dt_{i_k} }-\dfrac{dz_{j_k}(t_{i_k})}{dt_{j_k}}\right) \, \dfrac{1 }{z_{i_k}-z_{j_k}}\, G_{p}
 $$
 and finally, taking into account that $dz_{j_k}(t_{i_k})/dt_{j_k}=dz_{j_k}(t_{i_k})/dt_{i_k}$ we arrive to the following expression:
 \be
 \label{ki}
 \begin{array}{|c|}
 \hline\\
 \ \ \ \langle\, W(K)\,\rangle=\sum\limits_{n=0}^{\infty} \,\dfrac{\hbar^n}{(2\pi i)^n}\,\displaystyle\int\limits_{o(z_{1})<o(z_{2})<...o(z_{n})}\,\sum\limits_{p\,\in P_{2n}}\,(-1)^{p_\downarrow}\, \bigwedge\limits_{k=1}^{n}\, \dfrac{dz_{i_k}-dz_{j_k} }{z_{i_k}-z_{j_k}}\, G_{p}\ \ \ \\
 \\
 \hline
 \end{array}
 \ee
 The last expression (\ref{ki}) is the celebrated Kontsevich integral for knot $K$.

\subsection{Localization of Kontsevich Integral}
\subsubsection{Multiplicativity and braid representation}
The coefficients of Kontsevich integral (\ref{ki}) are given in terms of rather sophisticated meromorphic integrals. In this Section we describe the localization technique for computing the Kontsevich integral (KI) which gives a simpler combinatorial description of (\ref{ki}) \cite{ChDu}. The idea behind the localization is in fact very simple and based on miltiplicativity and factorization
of KI and uses braid representation of knots. Let us discuss these properties separately.

 \begin{wrapfigure}{l}{65mm}
  \vspace{-20pt}
  \begin{center}
    \includegraphics[scale=0.75]{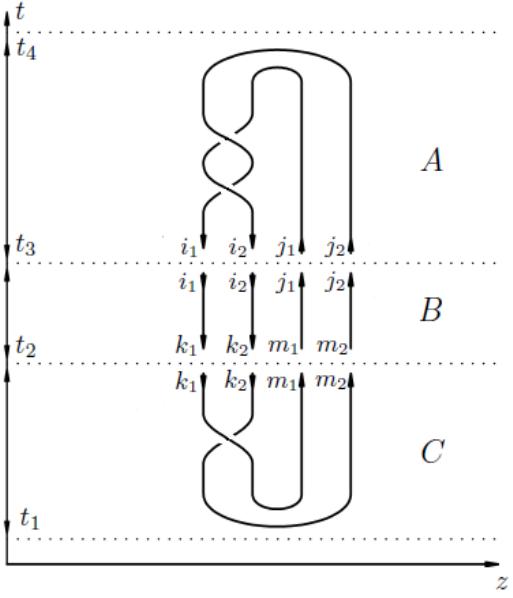}
  \end{center}
  \vspace{-25pt}
  \caption{Slices}
  \vspace{-10pt}
\label{fig:slices}
\end{wrapfigure}
The \textbf{multiplicativity} of KI means that we can cut the knot in the finite number of parts and compute KI for these parts separately, then KI for the whole knot can be computed as an appropriate product of these separate integrals. To illustrate the idea, let us consider knot $3_1$ (see \cite{katlas} for knot naming conventions) cut in three parts as in Figure \ref{fig:slices}. The Kontsevich integral computed for separate parts is not a number anymore but an operator of the form $V^{\otimes N_{in}}\rightarrow V^{\otimes N_{out}}$ where $N_{in}$ and $N_{out}$ are the numbers of incoming and outgoing lines correspondingly. For example let us cut the knot in three slices $(t_{4},t_{3})$, $(t_{3},t_{2})$ and $(t_{2},t_{1})$. Then the corresponding Kontsevich integrals are given by operators $A^{i_{1} i_{2} }_{j_{1} j_{2}}$,
$B_{ i_{1} i_{2} m_{1} m_{2} }^{ j_{1} j_{2} k_{1} k_{2} }$ and $C_{k_{1} k_{2} }^{m_{1} m_{2}}$  where we write their indices explicitly to emphasize that they are finite-dimensional tensors. Lower indices correspond to incoming lines and the upper to the outgoing ones.
The value of KI for the whole knot is the product of these tensors:
$$
\langle\, W(K)\,\rangle=\sum\limits_{i_{1} i_{2} j_{1} j_{2} \atop k_{1} k_{2} m_{1} m_{2}  }\,A^{i_{1} i_{2} }_{j_{1} j_{2}}\, B_{ i_{1} i_{2} m_{1} m_{2} }^{ j_{1} j_{2} k_{1} k_{2} }\, C_{k_{1} k_{2} }^{m_{1} m_{2}}
$$
For the entire knot KI is a number and not a tensor because the knot is closed and does not have any incoming or outgoing lines. Integral naturally brings us to the main idea of localization:
if we are able to represent the knots as a union of finite number of some special "fundamental" parts, then we have to compute KI for these parts only. In order to compute the KI, one needs to choose these special parts in a way such that the KI for them would have the most simple form.
Let us consider the middle slice $(t_{2},t_{3})$ in Figure \ref{fig:slices}.  Kontsevich integral $B$ for this part is obviously equal to identity, this is because this piece consists of four lines parallel to $t$ axis and the form $(dz_{i}-dz_{j})/(z_{i}-z_{j})$ in (\ref{ki}) vanishes as $dz=0$ on each vertical line:
$$
B=1^{\otimes 4} \ \  \textrm{or} \ \  B_{k_{1} k_{2} p_{1} p_{2}}^{n_{1} n_{2} m_{1} m_{2}}=\delta_{k_{1}}^{n_{1}}\delta_{k_{2}}^{n_2}\delta_{p_{1}}^{m_{1}}\delta_{p_2}^{m_{2}}
$$
Therefore, it is quite natural to use representations of knots "maximally extended" along the $t$ direction, such that the KI takes the simplest possible form. Such a representation is the well known braid representation of knots:

 \begin{wrapfigure}{r}{200pt}
  \vspace{-20pt}
  \begin{center}
    \includegraphics[scale=0.75]{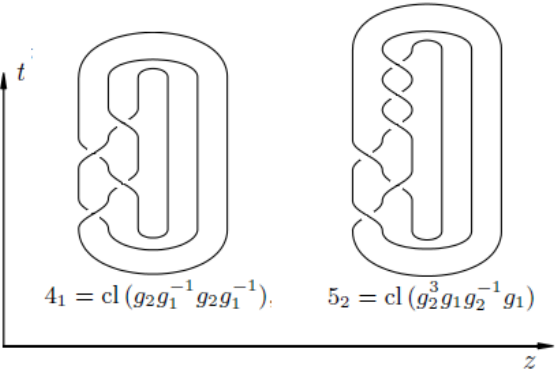}
  \end{center}
  \vspace{-25pt}
  \caption{Braid representation for knots}
\label{fig:knots}
  \vspace{-10pt}
\end{wrapfigure}
\textit{For any knot $K$ there is a number $n$ (not unique) such that it can be represented as a closure of some element (not unique) from braid group $\Br_{n}$.}

The meaning of this theorem is clear from the examples in Figure \ref{fig:knots}. Here the knots $4_1$ and $5_2$ are represented as closures of braids  $g_{2}g_{1}^{-1}g_{2}g_{1}^{-1}\in \Br_{3}$ and $g_{2}^{3}g_{1}g_{2}^{-1}g_{1}\in \Br_{3}$ correspondingly. The \textit{closure} is the operation that connects the top of the braid with its bottom stringwise. There exists a simple combinatorial algorithm to construct braid representation $b$ for any knot \cite{katlas}.

 \begin{wrapfigure}{r}{300pt}
  \vspace{-20pt}
  \begin{center}
    \includegraphics[scale=0.75]{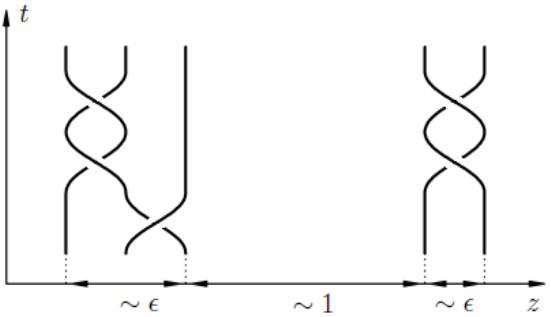}
  \end{center}
  \vspace{-20pt}
  \caption{Factorization}
\label{fig:fact}
  \vspace{-10pt}
\end{wrapfigure}
The last thing we need is the \textbf{factorization} of Kontsevich integral.

Let us introduce the distance between the strings in the braid. Then the following holds: if one can arrange the strings in the braid in a way that the distance between two groups of them is of order of different powers of some small parameter $\epsilon$, then these two groups give separate contributions to the KI.

More precisely:
Consider a braiding $b_{n}=b_{k}\otimes b_{n-k} \in \Br_{n}$. Let us assume that the sizes of $b_{k}$ and $b_{n-k}$ are much less ($O(\epsilon)$) than the distance between them. Then
$$
KI(b_{n})=KI(b_{k}) \otimes KI(b_{n-k}) + O(\epsilon)
$$
If $b_{n}=b_{k}\otimes b_{n-k}$ then the first $k$ strings of the braid do not cross the rest $n-k$ strings. For example in Figure \ref{fig:fact} the first three strings and the last two form two separate braids. If widths of these braids are ( $\sim \epsilon$) much less the distance between them ($\sim 1$), then the factorization theorem implies that the KI is the tensor product of two KI's for each separate braid.

\subsubsection{Choice of associators placement}

\begin{figure}[h] 
\begin{center}
\includegraphics[scale=0.75]{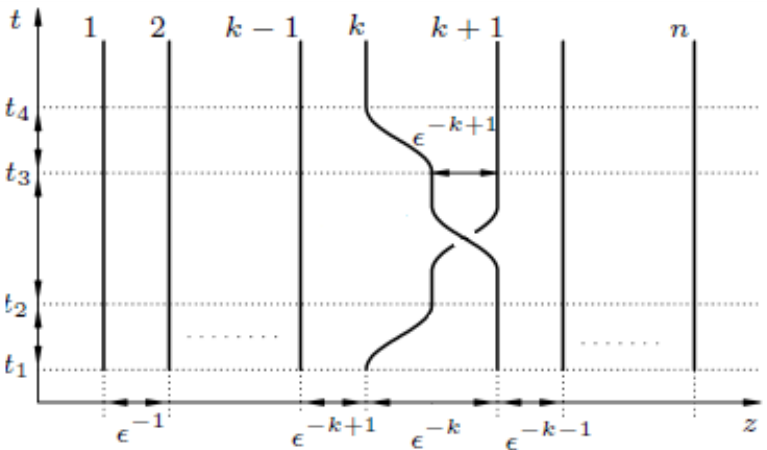}
\end{center}
  \caption{R-matrix with associators}
\label{fig:crossing}
\end{figure}

Now, with the help of these properties we are ready to describe the representation of a given knot for which KI takes the simplest form. Let knot $K$ be represented by closure of some braid $b\in \Br_{n}$, then we arrange the strings of the braid such that the distance between $k$-th and $k+1$-th strings is given by $\epsilon^{-k}$ where $\epsilon$ is some small formal parameter, Figure \ref{fig:crossing}:\\

\begin{wrapfigure}{r}{80pt}
  \vspace{-20pt}
  \begin{center}
    \includegraphics[scale=0.25]{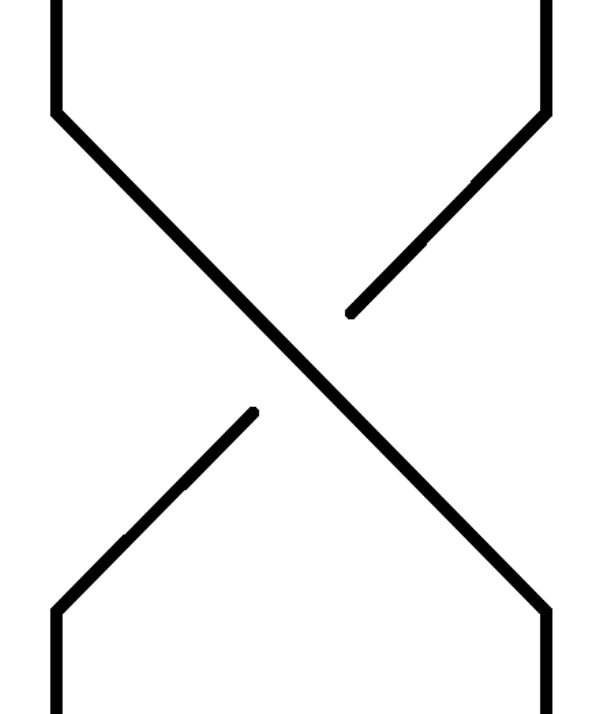}
  \end{center}
  \vspace{-20pt}
  \caption{R-matrix}
\label{fig:rmat}
  \vspace{-5pt}
\end{wrapfigure}
i.e. the distance between adjacent strings increases with the number of string. Suppose, that at some slice $(t_{4},t_{1})$ the braid $b$ has the crossing of $k$-th and $k+1$-th strings (Figure \ref{fig:crossing}). This crossing can be represented as a result of taking three consecutive steps:
\begin{enumerate}
\item In slice $(t_{4},t_{3})$ the $k$-th string goes closer to the string $k+1$, at the distance $\epsilon^{-k+1}$.
\item In slice $(t_{3},t_{2})$ two strings are crossed.
\item In slice $(t_{2},t_{1})$ the $k$-th string goes to its initial position at the distance $\epsilon^{-k}$ from $k+1$  string.
\end{enumerate}
If we arrange the strings in this way, then the distance between $k$-th and $k+1$-th strings at the interval $(t_{3},t_{2})$  is much less then the distance to any other string. Therefore, the factorization for KI gives the following contribution for slice $(t_{3},t_{2})$:
$$
KI_{(t_{3},t_{2})}=R_{k,k+1}^{\pm 1}=1_{1}\otimes...1_{k-1}\otimes R^{\pm 1} \otimes 1_{k+2} ...\otimes1_{n}
$$
\begin{wrapfigure}{r}{200pt}
  \vspace{-20pt}
  \begin{center}
    \includegraphics[scale=0.23]{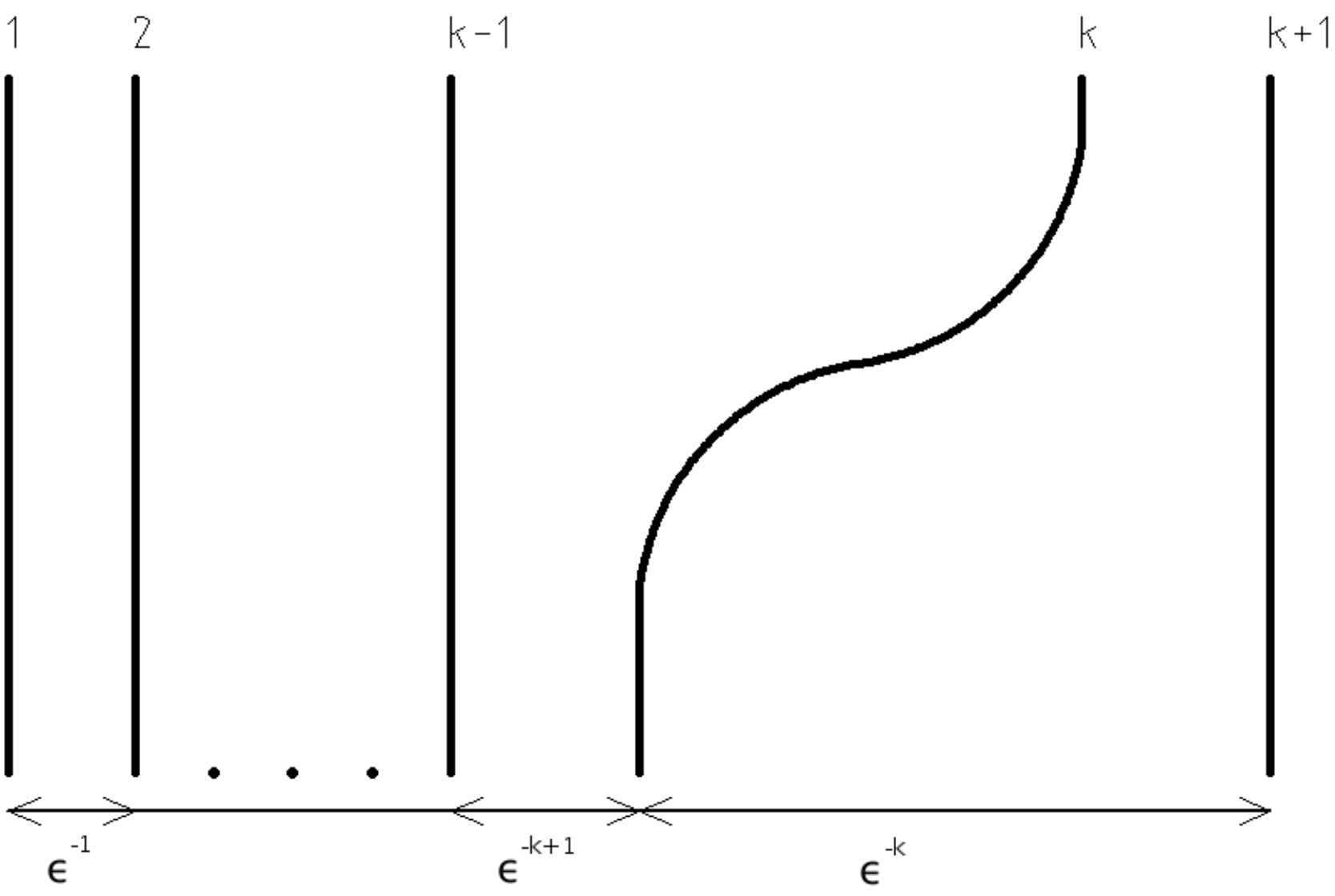}
  \end{center}
  \vspace{-20pt}
  \caption{Associator}
\label{fig:ass}
  \vspace{10pt}
\end{wrapfigure}
where $R^{\pm 1}$ is the so-called \textit{$R$-Matrix}, i.e. KI computed for the neighborhood of the crossing point which corresponds to the configuration of strings in Figure \ref{fig:rmat}; the sign depends on the orientation of the crossing.
Similarly, at slices ($t_{4},t_{3}$) and $(t_{2},t_{1})$ the distance between first and $k$-th string is much less the the distance to the other $n-k$ strings. Therefore, the factorization for the slice  ($t_{4},t_{3}$) gives:
$$
KI_{(t_{4},t_{3})}=\Psi_{1,2,...,k+1}=\Phi_{k+1} \otimes 1^{\otimes n-k-1}, \ \ \ KI_{(t_{2},t_{1})}=\Psi^{-1}_{1,2,...,k+1}=\Phi_{k}^{-1} \otimes 1^{\otimes n-k-1}
$$
where  $\Phi_{k+1}^{\pm 1}$ is the so-called \textit{associator}, given by KI for the configuration of strings represented in Figure \ref{fig:ass}. In summary, we see that the crossing of $k$-th and $k+1$-th strings in braid representation of the knot gives the contribution to the KI of the form of operator $R_{k,k+1}^{\pm}$ conjugated by operator $\Psi_{1,2,...,k+1}$:
$$
\begin{array}{|c|}
\hline\\
\ \ \ KI_{(t_{4},t_{1})}=\Psi_{1,2,...,k+1}\,R_{k,k+1}\,\Psi^{-1}_{1,2,...,k+1}\ \ \ \\
\\
\hline
\end{array}
$$
Denote this tensor by $X_k$:
\eql{xtens}{
X_k=\Psi_{1,2,...,k+1}\,R_{k,k+1}\,\Psi^{-1}_{1,2,...,k+1}.
}
\subsubsection{Formulas for R-Matrices and associators}
Explicit expressions for operators $R$ and $\Psi$ can be easily derived from computations of KI for the corresponding pieces of the braid representation of a knot. The KI for neighborhood of a crossing point gives:
\eql{rmat}{
R=\exp \Big(\frac{\hbar}{2\pi i} \Omega \Big), \ \ \ \Omega=T^{a}\otimes T^{a}
}
The calculation of $\Phi_{3}$ is explained in details in \cite{Le}, to describe it let us define:
$$
\Omega_{1 2}=\Omega \otimes 1, \ \ \ \Omega_{2 3}=1\otimes \Omega
$$
then we have:
\be
\label{e:assoc}
\Phi_{3}=1^{\otimes 3} + \sum\limits_{k=2}^{\infty} \Big(\dfrac{\hbar}{2\pi i}\Big)^{k} \sum\limits_{m \geq 0} \sum\limits_{{\textbf{p}>0\,\textbf{q}>0 \atop | \textbf{p} |+| \textbf{q} |=k} \atop l( \textbf{p} )=l( \textbf{q} )=m } (-1)^{|\textbf{q}|} \, \tau(p_1,q_1,...p_m,q_m)\times\nonumber\\
\sum\limits_{l(\textbf{r})=l(\textbf{s})=m \atop {0\leq \textbf{r} \leq \textbf{p},\, 0\leq \textbf{s}\leqq \textbf{q}}}\,(-1)^{|\textbf{r}|}\,\left( \prod\limits_{i=1}^{m} \Big( {{p_i}\atop{r_i}} \Big)\,\Big( {{q_i}\atop{s_i}} \Big) \right)\,\Omega_{23}^{|\textbf{s}|}\,\Omega_{1 2}^{p_1-r_1} \Omega_{2 3}^{q_1-s_1}...\Omega_{1 2}^{p_m-r_m} \Omega_{2 3}^{q_m-s_m}\Omega_{12}^{|\textbf{r}|}
\ee
Where $\textbf{p}=(p_1,p_2,...,p_m)$ is a vector with positive integer components. The length of the vector $l(\textbf{p})=m$ and $|\textbf{p}|=\sum p_{i}$. When we write $\textbf{p}>\textbf{q}$, it is understood as $p_{i}>q_{i}$ for all $i$, and $\textbf{p}>0$ means that $p_{i}>0$ for all $i$. The coefficients $\tau(p_1,q_1,...p_m,q_m)$ are expressed through multiple zeta functions as follows:
$$
\tau(p_1,q_1,...p_m,q_m)=\zeta(\underbrace{1,...,1}_{p_1-1},q_1+1,\underbrace{1,...,1}_{p_2-1},q_2+1,...,q_n+1)
$$
such that for example $\tau(1,2)=\zeta(3)$ and $\tau(2,1)=\zeta(1,2)$. The multiple zeta functions are defined as:
$$
\zeta(m_1,m_2,...,m_n)=\sum\limits_{0<k_{1}<k_{2}<...<k_{n}}\,k_{1}^{-m_1}k_{2}^{-m_2}...k_{n}^{-m_n} $$
Note, that $\tau(p_1,q_1,...p_m,q_m)=\tau(q_m,p_m,...q_1,p_1)$ so that, e.g. $\zeta(1,2)=\zeta(2)$ and $\zeta(1,3)=\zeta(4)$.
To define the associators $\Phi_{n}$ for $n>3$ we need the coproduct operator $\Delta: U(g)\rightarrow U^{\otimes 2}(g)$. Its action on the generators of the universal enveloping algebra $U(g)$ is defined as follows:
\be
\begin{array}{l}
\Delta(T^{a})=1\otimes T^{a}+T^{a}\otimes 1\\
\Delta(T^{a}T^{b})=1\otimes T^{a}T^{b}+ T^{a}\otimes T^{b}+T^{b}\otimes T^{a}+T^{a}T^{b}\otimes 1\\
....
\end{array}
\ee
With the help of this operator the higher associators are expressed through $\Phi_{3}$ with the help of the following recursive formula:
\be
\label{reqpr}
\Phi_{n+1}=\Delta\otimes 1^{\otimes(n-1)}\, \Phi_{n}
\ee
Formula (\ref{reqpr}) basically means that, e.g. to obtain $\Phi_4$ from $\Phi_3$ one needs to make the following substitutions in the formula for $\Phi_3$:
\begin{align*}
1\otimes T^{a}\otimes T^{a} &\longrightarrow 1\otimes 1\otimes T^{a}\otimes T^{a},\\
T^{a}\otimes 1\otimes T^{a} &\longrightarrow 1\otimes T^{a}\otimes T^{a}\otimes T^{a}+T^{a}\otimes 1\otimes T^{a}\otimes T^{a},\\
T^{a}T^{b}\otimes T^{a}\otimes T^{b} &\longrightarrow 1\otimes T^{a}T^{b}\otimes T^{a}\otimes T^{b}+T^{a}\otimes T^{b}\otimes T^{a}\otimes T^{b}+T^{b}\otimes T^{a}\otimes T^{a}\otimes T^{b}+T^{a}T^{b}\otimes 1\otimes T^{a}\otimes T^{b},\\
\dots &
\end{align*}
and so on. That is, one has to symmetrize the first tensor component of each term of $\Phi_3$ over the first and the second tensor components of $\Phi_4$.

One can derive (\ref{reqpr}) directly from Kontsevich integral consequentially for $n=3,4,...$, but it is simpler to note that in Figure \ref{fig:crossing} the first $k-1$ strings should give equivalent contribution to associator $\Phi_{k+1}$ because the distance between first and $k-1$-th strings is much less then the "width of associator" (the distance between the first and the $k+1$-th string). Therefore, operator $\Delta$ and the recursive procedure (\ref{reqpr}) have a clear physical meaning of "symmetrization" of contribution of the first $k-1$ strings to associator $\Phi_{k+1}$.
\newpage
\subsubsection{Caps}
 \begin{wrapfigure}{r}{200pt}
  \vspace{-20pt}
  \begin{center}
    \includegraphics[scale=0.75]{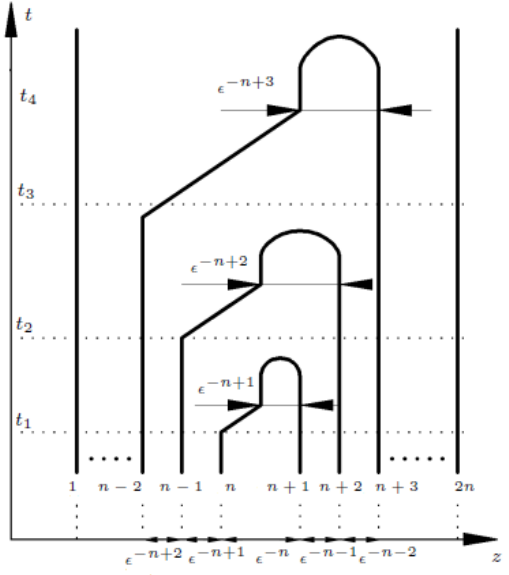}
  \end{center}
  \vspace{-20pt}
  \caption{Cap}
\label{fig:cap}
  \vspace{-15pt}
\end{wrapfigure}
To complete our consideration we should also find contributions to KI coming from the bottom and the top of the braid's closure. Fortunately, we do not need new operators as these contributions can be represented through already introduced operators $\Psi_{k}$.

Indeed, to make a closure of braid $b\in \Br_{n}$ we have to add to the braid $n$ straight strings, such that the total number of strings is $2n$ and then connect $(n-k)$-th string with the $n+k$-th one for all $k$ at the top and the bottom, for example as in Figure \ref{fig:cap}. In what follows we again imply that the distance between our $2 n$ strings increases with the number of string, such that the distance between $k$-th and $k+1$-th ones is of order $\epsilon^{-k}$. Consider the following procedure, first the $n$-th string goes closer to $n+1$ up to the distance $\epsilon^{-n}$ (slice $(t_1,t_2)$ in Figure \ref{fig:cap} ) then we connect them by a "hat" of width $\epsilon^{-n}$. The factorization theorem gives the contribution $\Psi^{-1}_{1,2...,n+1}$ for this slice (due to the hat contribution being trivial). In the slice $(t_2,t_3)$ we have $2n-2$ strings and we can iterate the procedure one more time which will give the contribution of the form $\Psi^{-1}_{1,2,...n-1,n+2}$. Finally, the whole top closure gives:
\eql{ttens}{
T_{n} = \Psi_{1,2,2n-1}^{-1} \Psi_{1,2,...n-1,n+2}^{-1}... \Psi_{1,2,...n+1}^{-1}.
}
It is assumed here that the indices corresponding to strings which terminate in a cap are contracted.

Analogously for the bottom of the closure:
\eql{btens}{
B_{n} = \Psi_{1,2,2n-1} \Psi_{1,2,...,n-1,n+2}... \Psi_{1,2,...,n+1}
}
Again, one should contract all the indices corresponding to caps.

\subsubsection{General combinatorial formula for Kontsevich integral}
Now we know the contributions of all parts of the braid closure. In order to write down the answer, let us introduce the symbol $\mathop{\prod}^{\rightarrow}$ representing \textit{the ordered product}.

The answer is then as follows.
 
 \textit{Let knot $K$ be represented as the closure of a braid $b \in \Br_{n}$:
$$
b=\prod\limits_{k}^{\rightarrow} g_{k}
$$
then the KI for the knot is given by the following expression\footnote{One should understand that the symbol "$\tr$" in (\ref{e:KIF}) stands for the contraction of the tensors $X_{i}$, $B_{n}$ and $T_{n}$ corresponding to their position in the braid.}:
}
\be
\label{e:KIF}
\begin{array}{|c|}
\hline\\
\ \ \ KI\Big(\,\prod\limits_{k}^{\rightarrow}\, b_{i_{k}}\,\Big) = \tr\Big(\, T_{n} \prod\limits_{k}^{\rightarrow}\, X_{i_{k}} B_{n}\,\Big)\ \ \ \\
\\
\hline
\end{array}
\ee
where the tensors $T_n$, $X_i$ and $B_n$ are given by formulas \re{ttens}, \re{xtens} and \re{btens} correspondingly.

\subsection{Technique of computation}

In order to finally compute the coefficients of Kontsevich integral, one needs to take the following steps.

First of all, note that basic elements of formula \re{KIF} are contractions $T^aT^a$ where one $T^a$ stands in some tensor component and the other in another one. Note that one can represent it as chord drawn on the knot from the place on the knot corresponding to the first $T^a$ to the place corresponding to the second $T^a$. Then, in order to compute the given order of \re{KIF}, one has to consider all of the terms in \re{KIF} of that order, then for every such term draw a chord corresponding to every contracted pair of generators and untie the knot obtaining a \textit{chord diagram}.

More precisely, let $n$ be a given order in $\hbar$ which we want to compute. In this order, Kontsevich integral is a linear combination of group factors $G_{p}=\tr\Big( T^{\sigma_{p}(1)} T^{\sigma_{p}(2)}...T^{\sigma_{p}(2n)} \Big)$ corresponding to pairings $p=\Big((i_1,j_1) (i_2,j_2)...(i_n,j_n)\Big)$.

One may associate with every such group factor a \textit{chord diagram}, which corresponds to the particular pairing in a straightforward way. For example, there is the following correspondance: 
$$
\tr\br{T^aT^bT^cT^aT^cT^b} \leftrightarrow \chdiag
$$

Thus, Kontsevich integral in order $n$ is a linear combination of chord diagrams with $n$ chords.

In these terms, formal formula \re{KIF} may be expressed as the following collection of steps.

Let there be associators $\Psi_1,\dots,\Psi_k$ and R-matrices $R_1,\dots,R_l$ assigned to a given knot $K$. Then, to obtain Kontsevich integral in order $n$ one has to take the sum over all ordered partitions of $n$ into $k+l$ parts:
\eq{
n=\phi_1+\dots+\phi_k+r_1+\dots+r_l,
}
where all $\phi_i$ and $r_i$ are nonnegative integers.

The R-matrix part is easier, for every R-matrix $R_i$ we just insert $r_i$ consequential chords (propagators) in the corresponding place on the knot, and take the coefficient $1/r_i!$.

The associator part is trickier, there we have several summands due to the form of formula \re{assoc} for the associator, and to the fact that we should include also braids to the left, as it is done in formula (\ref{reqpr}). For every $\Psi_i$ therefore there is a whole sum of ways to insert $\phi_i$ chords in the knot, with corresponding coefficients. That means that in a given order there are several terms of this order in expression \re{assoc}. This number then multiplies even further when one takes into account (\ref{reqpr}).

Then, for given partition $n=\phi_1+\dots+\phi_k+r_1+\dots+r_l$, we take the product of all described above parts corresponding to R-matrices and associators. After expanding the brackets, we obtain a sum of several ways to insert $n$ chords in the knot with corresponding coefficients.

The only thing left is to untie the knot into a circle thus obtaining a linear combination of chord diagrams with $n$ chords. Algorithmically, it can be done in a very straightforward way just by assigning an ordinal number to each part of the knot which belongs to an associator or to an R-matrix.

Then we sum over all partitions of $n$ and obtain the coefficients of order $n$ in Kontsevich integral.

Refer to Appendix \ref{app:fek} for more detailed description of the computation of KI coefficients for the figure-eight knot, i.e. knot $4_1$.

\section{From Kontsevich integral to Vassiliev invariants}
\label{numvas}
\subsection{Chern-Simons definition of Vassiliev invariants}
\label{csvi}

\begin{figure}[!ht]
\centering\leavevmode
\includegraphics[scale=0.75]{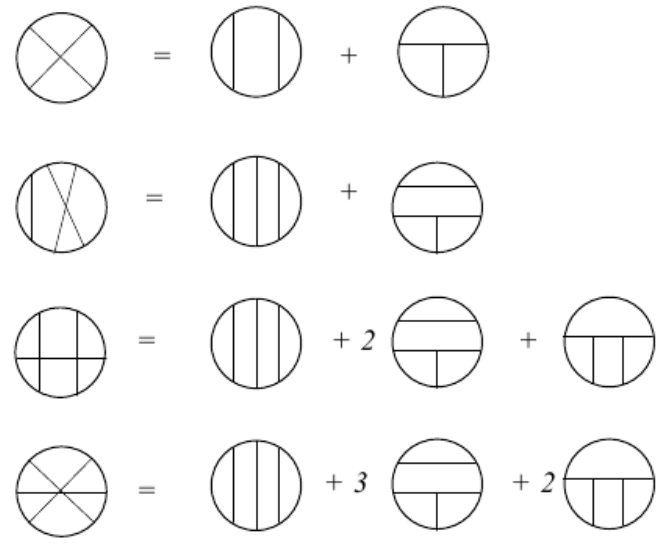}
\caption{Relation between trivalent diagrams and chord diagrams up to order 3}
\label{p_triv}
\end{figure}

\begin{figure}[h]
\centering\leavevmode
\includegraphics[width=10 cm]{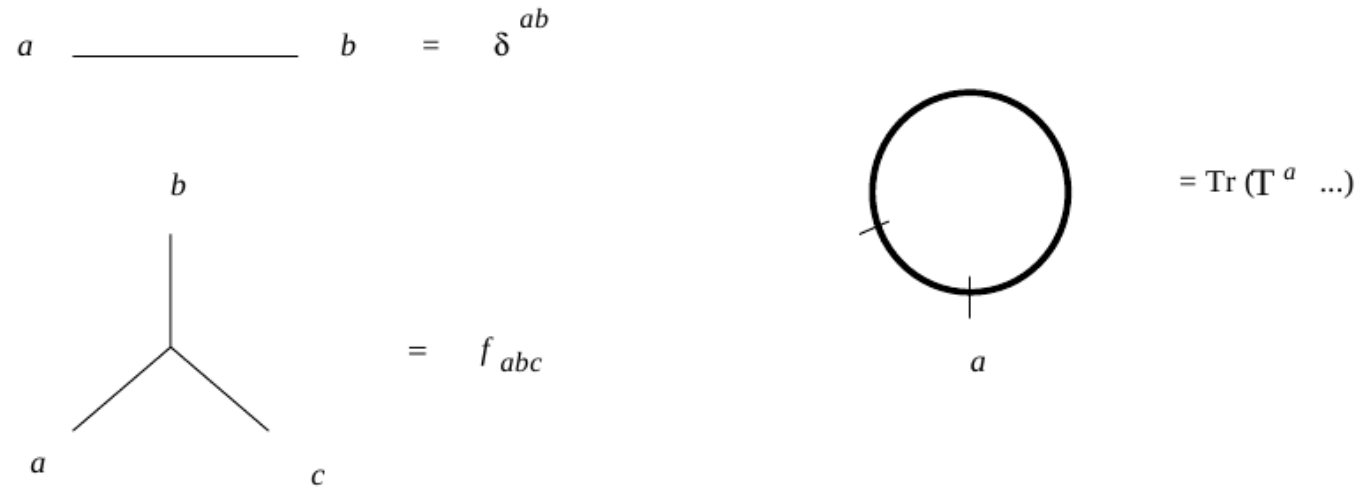}
\caption{Group-theoretical rules}
\label{rules}
\end{figure}

As discussed above, the perturbation theory for CS leads to the following $\hbar$ expansion of vev for Wilson loops (\ref{eq1}): 
\eq{ \label{A}
<W_{R}(K)>\,=\sum\limits_{n=0}^{\infty}\,\hbar^n\,\sum_{n=1}^{\dim(H_{n})}\,V_{n,m}\,G_{n,m}
} 
here, $K$ is the knot represented by the Wilson loop, $R$ is the representation living on the loop, $V_{k,m}$-are some rational coefficients depending on the knot and $G_{k,m}$ are the group factors called the chord diagrams with $n$ chords. The chord diagrams with $n$ chords form the vector space $H_{n}$. The
coefficients $V_{n,m}$ are not knot invariants because the chord diagrams are not independent, and to express this sum through invariants we have to choose some basis in $H_{n}$. The dimensions of $H_{n}$ are summarized in the table:
\begin{equation}
\begin{array}{|c|c|c|c|c|c|c|}
\hline
n&1&2&3&4&5&6\\
\hline
dim(H_{n})&1&1&1&3&4&9\\
\hline
\end{array}
\end{equation}
In order to pass to Vassiliev invariants we have to choose some basis in the space of chord diagrams. We do it following \cite{La5}, refer to that paper for details. The so-called trivalent diagrams are introduced in a way represented for orders two and three in Figure \ref{p_triv}. Group-theoretical rules for graphical representation of chords and trivalent diagrams are presented in Figure \ref{rules}. For the general definition of trivalent diagrams refer to \cite{La5}, see also \cite{La6}.  

Let us explain the definition of trivalent diagrams on the first relation from Figure \ref{p_triv}: $$T^aT^bT^cT^d\delta^{ac}\delta^{bd}=T^aT^bT^aT^b=\mathpic{0.4}{\draw[black] (0,0) circle (1); \draw[-,black,thick] (-0.707106,0.707106) -- (0.707106,-0.707106); \draw[-,black,thick] (0.707106,0.707106) -- (-0.707106,-0.707106);}$$ 
$T^aT^bT^cT^d\delta^{ad}\delta^{bc}=T^aT^bT^bT^a=\mathpic{0.4}{\draw[black] (0,0) circle (1); \draw[-,black,thick] (-0.5,0.866025) -- (-0.5,-0.866025); \draw[-,black,thick] (0.5,0.866025) -- (0.5,-0.866025);}=T^aT^bT^aT^b-T^aT^bT^aT^b+T^aT^bT^bT^a=T^aT^bT^aT^b-T^aT^b\left(T^aT^b-T^bT^a\right)=T^aT^bT^aT^b-T^aT^b\left[T^aT^b\right]= T^aT^bT^aT^b-f^{abc}T^aT^bT^c=\mathpic{0.4}{\draw[black] (0,0) circle (1); \draw[-,black,thick] (-0.707106,0.707106) -- (0.707106,-0.707106); \draw[-,black,thick] (0.707106,0.707106) -- (-0.707106,-0.707106);} - \mathpic{0.4}{\draw[black] (0,0) circle (1); \draw[-,black,thick] (-1,0) -- (1,0); \draw[-,black,thick] (0,0) -- (0,-1);}$

\begin{figure}[h!]
\centering\leavevmode
\includegraphics[width=13 cm]{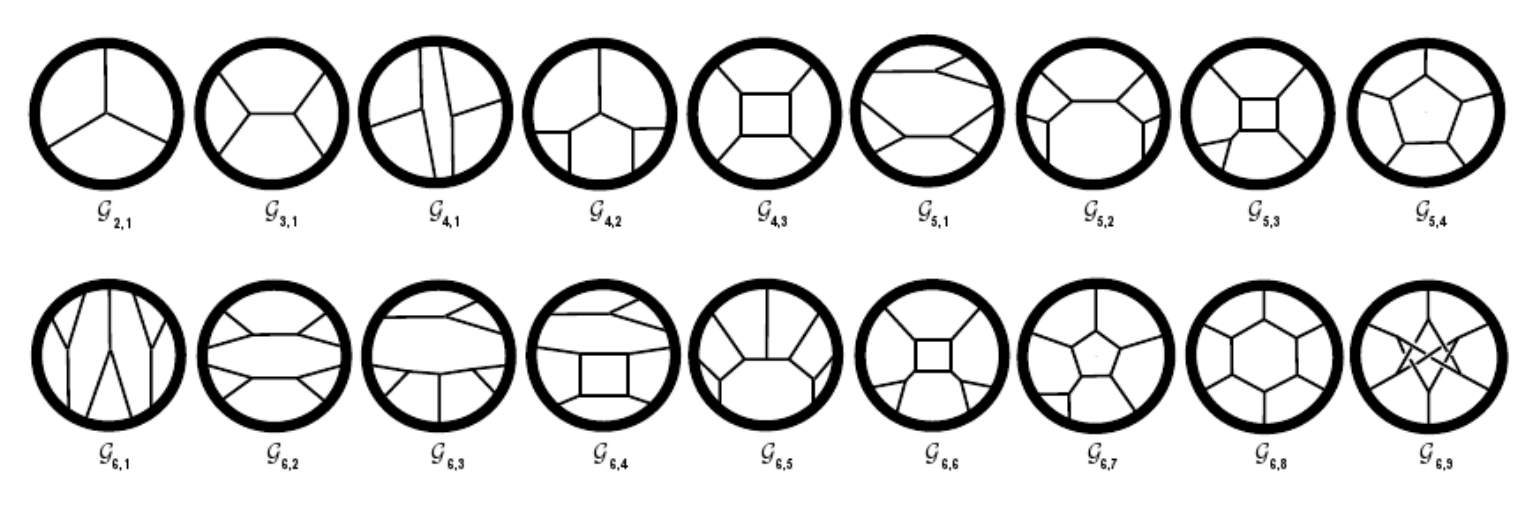}
\caption{Trivalent diagrams}
\label{p_chords}
\end{figure}

In Figure \ref{p_chords} one can find a collection of trivalent diagrams that form the so-called \textit{canonical basis} $\{{\cal{G}}_{ij} \}$ of $H_{n}$ up to order six.

Using this basis we rewrite (\ref{A}) through invariants:
\eq{ \label{B}
<W_{R}(K)>\,=\sum\limits_{n=0}^{\infty}\,\hbar^n\,\sum_{m=1}^{\dim(H_{n})}\,{\cal{V}}_{n,m}\,{\cal{G}}_{n,m}
}
Here ${\cal{V}}_{ij}$ are the so called finite-type or Vassiliev invariants of knots. They depend only on the knot under consideration but not on the group and its representation.

Now let us introduce the primitive Vassiliev invariants.
It is a well known fact that the expansion of logarithm of any correlator in any QFT contains only connected Feynman diagrams (for more details about this situation in the Chern-Simons perturbation theory see \cite{La8}). This fact immediately leads to the following summation of
\eq{ 
\label{C}
<W_{R}(K)>\,=\prod\limits_{n=0}^{\infty}\,\prod_{m=1}^{{\cal{N}}_n}\,
\exp\left(  \hbar^n{\cal{V}}^c_{n,m}\,{\cal{G}}^c_{n,m} \right),
}
where ${\cal{G}}^c$ are connected diagrams, ${\cal{V}}^c$ are primitive Vassiliev invariants. Here ${\cal{N}}_n$ is dimension of the space of connected chord diagrams (or equivalently the space of primitive Vassiliev invariants). The dimensions of these spaces up to order 6 are given in the following table:
\begin{equation}
\begin{array}{|c|c|c|c|c|c|c|}
\hline
n&1&2&3&4&5&6\\
\hline
{\cal{N}}_n&1&1&1&2&3&5\\
\hline
\end{array}
\end{equation}
The meaning of the expression (\ref{C}) is that ${\cal{V}}_{i,j}$ in (\ref{B}) are not independent. In fact only those coefficients ${\cal{V}}_{ij}$ are independent, for which the corresponding diagram ${\cal{G}}_{ij}$ is connected. Comparing $\hbar$ expansion of (\ref{C}) with (\ref{B}) we, for example, find:
$$
{\cal{V}}_{4,1}=1/2\, {\cal{V}}_{2,1}^2
$$
And finally: \textit{The Vassiliev invariants form a graded ring freely generated by primitive invariants. }

\subsection{Numerical results for invariants up to order 6 and families of knots}
\label{NumRes}
Now one can use computational technique from the previous Section \ref{KIP} and results from subSection \ref{csvi} to calculate the Kontsevich integral. In this way we calculate Vassiliev invariants up to level 6 inclusive for knots with number of self-interSections up to 14 inclusive. These results are available in \cite{katlas}. Here we list just a few examples:
\begin{small}
\begin{equation}
\begin{array}{|c|c|c|c|c|c|c|c|c|c|c|c|c|c|c|c|c|c|c|}
\hline
&{\cal{V}}_{2,1}&{\cal{V}}_{3,1}&{\cal{V}}_{4,1}&{\cal{V}}_{4,2}&{\cal{V}}_{4,3}&{\cal{V}}_{5,1}&{\cal{V}}_{5,2}&{\cal{V}}_{5,3}&{\cal{V}}_{5,4}&{\cal{V}}_{6,1}&{\cal{V}}_{6,2}&{\cal{V}}_{6,3}&{\cal{V}}_{6,4}&{\cal{V}}_{6,5}&{\cal{V}}_{6,6}&{\cal{V}}_{6,7}&{\cal{V}}_{6,8}&{\cal{V}}_{6,9}\\
\hline
3_1&4 &	−8 &	8 	&\frac{62}{3} &	\frac{10}{3} &	−32 &	-\frac{176}{3}& 	-\frac{32}{3} 	&−8 &	\frac{32}{3} &	32 &	\frac{248}{3}& 	\frac{40}{3} &	\frac{5071}{30} &	\frac{58}{15}& 	\frac{3062}{45} 	&\frac{17}{18}& 	\frac{271}{30}\\
\hline
4_1&−4 &	0 &	8 	&\frac{34}{3} &	\frac{14}{3} &	0 	&0 &	0 &	0 	&-\frac{32}{3}& 	0 	&-\frac{136}{3} &	-\frac{56}{3} 	&-\frac{1231}{30} &	\frac{142}{15} 	&-\frac{1742}{45} 	&\frac{79}{18} 	&-\frac{271}{30}\\
\hline
5_1&12&	−40&	72&	174&	26&	−480&	-\frac{2512}{3}&	-\frac{448}{3}&	−104&	288&	800&	2088&	312&	\frac{41151}{10}&	\frac{2494}{15}&	\frac{7634}{5}&	\frac{43}{2}&	\frac{1951}{10}\\
\hline 
5_2&8&	−24&	32&	\frac{268}{3}&	\frac{44}{3}&	−192&	−368&	−64&	−56&	\frac{256}{3}&	288&	\frac{2144}{3}&	\frac{352}{3}&	\frac{22951}{15}&	-\frac{28}{5}&	\frac{29764}{45}&	\frac{137}{9}&	\frac{1351}{15}\\
\hline
6_1&−8&	8&	32&	\frac{116}{3}&	\frac{52}{3}&	−64&	-\frac{304}{3}&	-\frac{64}{3}&	−24&	-\frac{256}{3}&	32&	-\frac{928}{3}&	-\frac{416}{3}&	-\frac{2791}{15}&	\frac{884}{15}&	-\frac{10084}{45}&	\frac{343}{9}&	-\frac{871}{15}\\
\hline
6_2&−4&	8&	8&	\frac{34}{3}&	\frac{38}{3}&	−32&	-\frac{208}{3}&	-\frac{64}{3}&	−24&	-\frac{32}{3}&	32&	-\frac{136}{3}&	-\frac{152}{3}&	\frac{2129}{30}&	\frac{662}{15}&	-\frac{1862}{45}&	\frac{463}{18}&	-\frac{751}{30}\\
\hline
6_3&4&	0&	8&	\frac{14}{3}&	-\frac{14}{3}&	0&	0&	0&	0&	\frac{32}{3}&	0&	\frac{56}{3}&	-\frac{56}{3}&	\frac{511}{30}&	\frac{418}{15}&	-\frac{1858}{45}&	\frac{65}{18}&	-\frac{449}{30}\\
\hline
\end{array}
\end{equation}              \end{small}
There is also one more technique to compute these invariants. It was suggested by Alvarez and Labastida in \cite{La9}. We use this technique to check our results.

\subsubsection{Willerton's fish and families of knots}
On 18 Vassiliev invariant up to order six
\eq{
{\cal{V}}_{2,1},\,{\cal{V}}_{3,1},\,{\cal{V}}_{4,1},\,{\cal{V}}_{4,2},\,{\cal{V}}_{4,3},\,{\cal{V}}_{5,1},\,{\cal{V}}_{5,2},\,{\cal{V}}_{5,3},\,{\cal{V}}_{5,4},\,{\cal{V}}_{6,1},\,{\cal{V}}_{6,2},\,{\cal{V}}_{6,3},\,{\cal{V}}_{6,4},\,{\cal{V}}_{6,5},\,{\cal{V}}_{6,6},\,{\cal{V}}_{6,7},\,{\cal{V}}_{6,8},\,{\cal{V}}_{6,9}
}
there are 6 relations
\eqs{
\label{e:vas_rel}
{\cal{V}}_{4,1}=&1/2\, {\cal{V}}_{2,1}^2, \\
{\cal{V}}_{5,1}=&{\cal{V}}_{2,1}{\cal{V}}_{3,1}, \\
{\cal{V}}_{6,1}=&{1\over 6}{\cal{V}}_{2,1}^3,\\
{\cal{V}}_{6,2}=&1/2\, {\cal{V}}_{3,1}^2, \\
{\cal{V}}_{6,3}=&{\cal{V}}_{2,1}{\cal{V}}_{4,2}, \\
{\cal{V}}_{6,4}=&{\cal{V}}_{2,1}{\cal{V}}_{4,3}. \\
}
Remaining 12 Vassiliev invariants are generally thought to be independent.

However, if one plots ${\cal{V}}_{3,1}$ against ${\cal{V}}_{2,1}$ for all knots of given number of crossings, one obtains filled region with form resembling fish, which was discovered by Willerton in \cite{Wil} and is hence called Willerton's fish, see Figure \ref{p_fish}. See Appendix \ref{apics} where we present some new even more peculiar figures representing relations between Vassiliev invariants.

The boundaries of the fish can be fitted with cubic polynomial. This suggests, that there are some additional relations on Vassiliev invarants. Note also, that for prime knots, which are actually plotted in Figure \ref{p_fish} coordinates are ``quantized'' -- they are integer multiples of ${\cal{V}}'s$ for the trefoil. For higher Vassiliev invariants in our chosen basis it is not the case.

\begin{figure}[h!]
\centering\leavevmode
\includegraphics[width=10 cm]{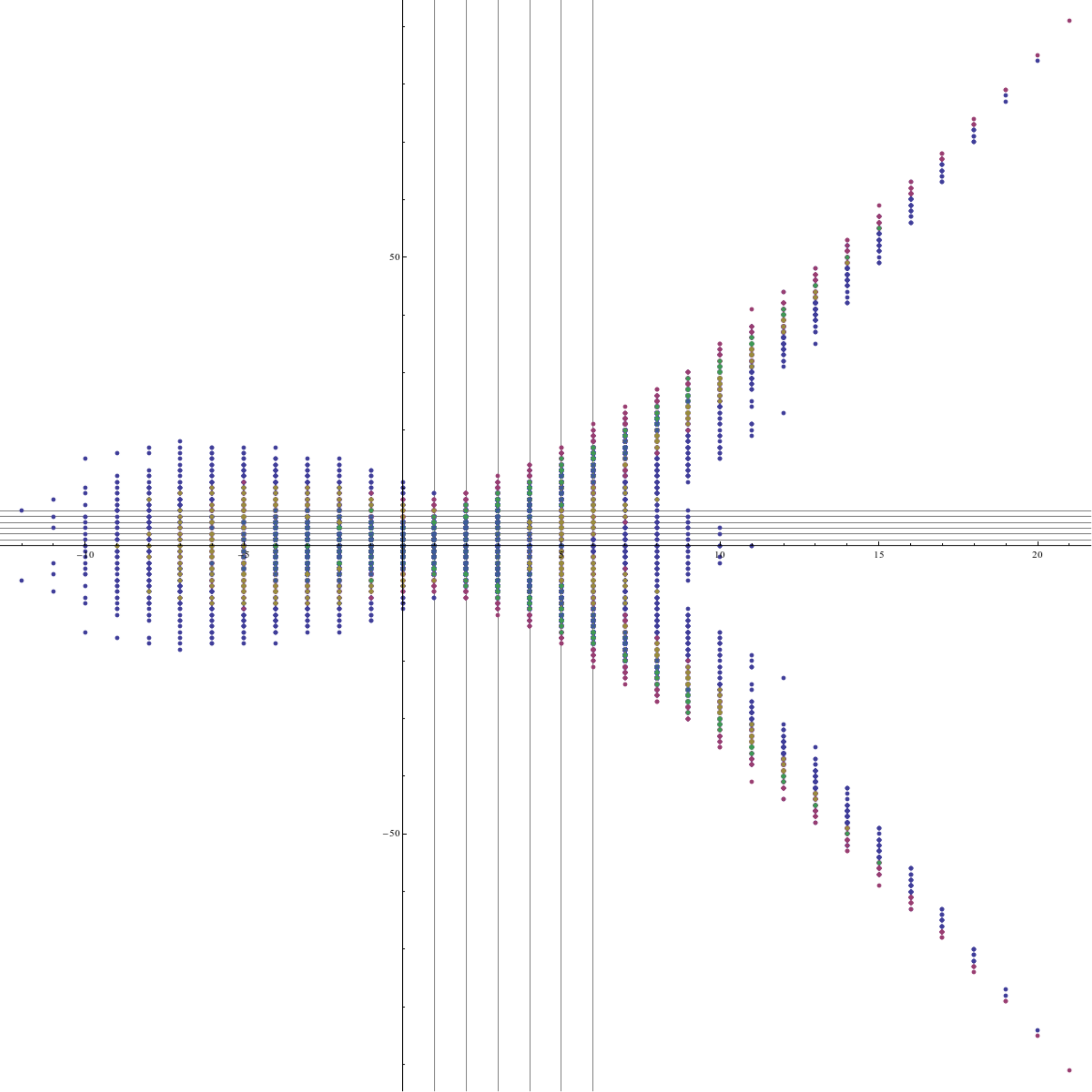}
\caption{Willerton's fish, i.e. ${\cal{V}}_{3,1}$ plotted against ${\cal{V}}_{2,1}$ for knots with up to 14 crossings}
\label{p_fish}
\end{figure}

Thus, it is natural to search for relations on Vassiliev invariants up to 6 order, i.e. to search for vanishing polynomials in ${\cal{V}}$ of order up to 6. If such polynomials can be found it would mean that Vassiliev invariants are not independent. However, it turned out, that there are no such vanishing expressions.

Nevertheless, the following expression vanishes for all knots up to 6 crossings and for quite big number of knots with more crossings, for example, for 92 out of 165 knots with 10 crossings:
\eqlm{constr}{
F_1=1/384\bigg(\frac{16 {\cal{V}}_{2,1}}{15}+2 {\cal{V}}_{2,1}^2+{\cal{V}}_{2,1}^3-6 {\cal{V}}_{3,1}-6 {\cal{V}}_{2,1} {\cal{V}}_{3,1}-3 {\cal{V}}_{3,1}^2-4 {\cal{V}}_{4,2}-6 {\cal{V}}_{2,1} {\cal{V}}_{4,2}+\\+4 {\cal{V}}_{4,3}+6 {\cal{V}}_{2,1} {\cal{V}}_{4,3}+6 {\cal{V}}_{5,2}-6 {\cal{V}}_{5,3}-6 {\cal{V}}_{5,4}+6 {\cal{V}}_{6,5}-6 {\cal{V}}_{6,6}-6 {\cal{V}}_{6,7}-6 {\cal{V}}_{6,8}\bigg).
}
This is very interesting, because the general polynomial of order 6 has 17 free coefficients, while polynomial \re{constr} vanishes on 92 knots among solely the knots with 10 crossings!

Another numerical experimental fact is that the value of \re{constr} is generally close to the value of \textit{writhe number} (see Section \ref{TG} for the definition), and tends to coincide with it, but not for all knots.

\section{Temporal gauge}
\label{TG}
In Section \ref{kics} we declare that from Chern-Simons theory in Lorentz gauge one can get integral representations for Vassiliev invariants (\ref{gi1}, \ref{gi2}). Despite beautiful the integral representation for Vassiliev invariants is too complicated to compute and it is difficult to use them for investigations. From the other side, the knots are simple combinatorial objects and it is not surprisingly that simpler combinatorial formulas for the invariants should exist. In this Section we are going to define combinatorial objects related to the knots called \textit{writhe numbers}. It will be shown that writhe numbers can be derived from the Wilson loop operator in temporal gauge. However it turns out that it is impossible to express Vassiliev invariants through such writhe numbers. Because of this reason we define \textit{colored writhe numbers} in a combinatorial way by analogy with the combinatorial definition of writhe numbers, because temporal gauge consideration does not lead us to them. Then it is possible to express Vassiliev invariants through colored writhe number up to order 4 at least (this fact was found in \cite{PV1, PV2}). 

\subsection{Writhe numbers from temporal gauge}
\begin{figure}[h!]
\centering\leavevmode
\includegraphics[width=4 cm]{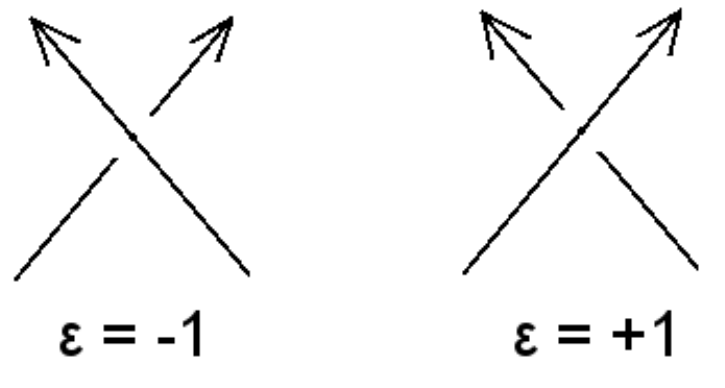}
\caption{Crossing signs}
\label{cr_signs}
\end{figure}

Consider Chern-Simons action in the temporal gauge $A_0 = 0$. Then the propagator takes the following form:
\eqs{\label{temprop1}
\langle A_0^a(x), A_{\mu}^b(y)\rangle &= 0, \hspace{5mm} \mu=0,1,2, \\
\langle A_{\mu}^a(x), A_{\nu}^b(y)\rangle &= \dfrac{1}{2}\varepsilon^{\mu\nu}\delta^{ab}\delta(x_1-y_1)\delta(x_2-y_2)\rm{sign}(x_0 - y_0), \hspace{5mm} \mu=0,1,2.
\label{temprop2}
}
Now let us consider the vacuum expectation value of the Wilson loop operator (\ref{eq1}):
\eq{
\langle W(K)\rangle\ = \sum\limits_{n=0}^{\infty} \oint dx_{1}\int dx_{2}...\int dx_{n} \langle A^{a_1}(x_{1})A^{a_2}(x_{2})...A^{a_3}(x_{n})\rangle \tr(T^{a_1} T^{a_2}...T^{a_n}).
}
Taking into account the propagators (\ref{temprop1}), (\ref{temprop2}) consider term of vev with $n=2$ in details:
\eq{
\int dx_{\mu} \int dy_{\nu} \langle A(x_{\mu})A(x_{\nu}))\rangle = \dfrac{1}{2} \int \int dx_{\mu}dy_{\nu}\varepsilon^{\mu\nu}\delta^{ab}\delta(x_1-y_1) \delta(x_2-y_2)\rm{sign}(x_0 - y_0).
}
Let us parametrize the knot by a parameter $t$ running from 0 to 1, then we can rewrite the last integral in the following form:
\eq{
\int_0^1\int_0^1 dt_1dt_2\br{\dfrac{dx_1}{dt_1}\dfrac{dy_2}{dt_2} - \dfrac{dx_2}{dt_1}\dfrac{dy_1}{dt_2}} \delta\br{x_1(t_1) - y_1(t_2)} \delta\br{x_2(t_1) - y_2(t_2)} \rm{sign}\br{x_0(t_1) - y_0(t_2)}
}
To perform the integration we need to solve the following
equations:
\eql{Last}{
 \left\{\begin{array}{c}
x_{1}( t_{1} )-y_{1}(t_{2})=0\\
\\
x_{2}( t_{1} )-y_{2}(t_{2})=0
\end{array}\right.
}
The solutions of these equations are the self-interSection points of two-dimensional curve $(x_{1}(t),x_{2}(t))$ which is the projection of the knot $c$ on the plane $(x_{1},x_{2})$. Let us denote by $t_{1}^{k}<t_{2}^{k}$ the values of the parameter $t$ in the interSection points, then the two-dimensional delta-function
in the integral can be represented in the following form:
$$
\delta(x_{1}(t_{1})-y_{1}(t_{2}))\,\delta(x_{2}(t_{1})-y_{2}(t_{2}))=\sum\limits_{k}
\dfrac{\left(
\delta(t_{1}-t_{1}^k)\delta(t_{2}-t_{2}^k)+\delta(t_{1}-t_{2}^k)\delta(t_{2}-t_{1}^k)
\right)}{|\dfrac{dx_{1}}{dt_{1}}\,\dfrac{dy_{2}}{dt_{2}}-
\dfrac{dx_{2}}{dt_{1}}\,\dfrac{dy_{1}}{dt_{2}}|}
$$
Substituting this expression into \re{Last} and integrating over $t_{1}$ and $t_{2}$ we arrive to the following simple
expression:
\begin{equation}
\int dx_{\mu} \int dy_{\nu} \langle A(x_{\mu})A(x_{\nu}))\rangle=\sum_{k} \epsilon_{k},
\end{equation}
where the quantities $\epsilon_{k}$ are the "sings" of the interSection points. They can take values $\pm 1$ and are defined in the following way:
\begin{equation}
\label{signs}
\epsilon_{k}=\dfrac{\dfrac{dx_{1}}{dt_{1}}(t_{1}^{k})\,\dfrac{dy_{2}}{dt_{2}}(t_{2}^{k})-
\dfrac{dx_{2}}{dt_{1}}(t_{1}^{k})\,\dfrac{dy_{1}}{dt_{2}}(t_{2}^{k})}{|\dfrac{dx_{1}}{dt_{1}}(t_{1}^{k})\,\dfrac{dy_{2}}{dt_{2}}(t_{2}^{k})-
\dfrac{dx_{2}}{dt_{1}}(t_{1}^{k})\,\dfrac{dy_{1}}{dt_{2}}(t_{2}^{k})|}\,\textrm{sign}(x_{0}(t_{1}^{k})-
y_{0}(t_{2}^{k}))
\end{equation}
Thus, Chern-Simons theory in temporal gauge naturally leads us to the following notion.
\dfn{
Writhe number for oriented knot $w(\K)$ is defined as the sum of the signs of all the crossings:
\eq{
w_1(\K) := \summ{p}{} \ep(p),
}
}
where the crossing signs are $+1$ or $-1$ as indicated in Figure \ref{cr_signs}.

\begin{figure}[h!]
\centering\leavevmode
\includegraphics[width=3 cm]{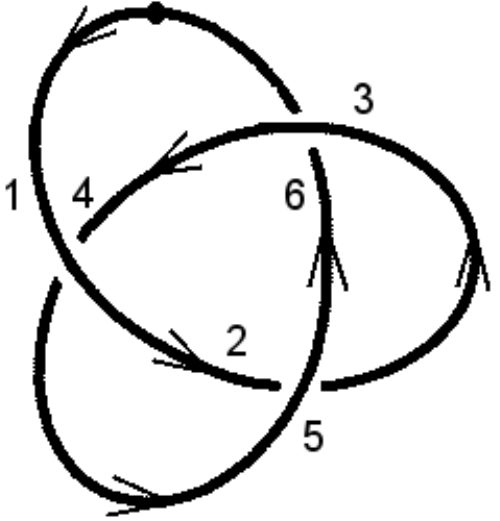}
\caption{Trefoil}
\label{trefoil}
\end{figure}

Writhe numbers can be represented graphically, if one gives another equivalent definition. Let us choose the origin on the knot and the orientation. When we are going along the knot, we meet every crossing point twice. We enumerate all crossing points by increasing sequence of the natural numbers. Then every crossing point is defined by the pair of numbers $(i_1,i_2), \ i_1 \neq i_2$ (see, for example, Figure \ref{trefoil}).
So, we get the following different definition of the writhe number.
\dfn{
Writhe number for oriented knot $w(\K)$ is defined as a sum of the crossing signs:
\eq{
w_1(\K) := \summ{i_1<i_2}{} \ep_{i_1i_2}.
}
}
Now we can depict the writhe number by a chord diagram, which is called the Gauss diagram:
\eq{
w_1(\K) = \summ{i_1<i_2}{} \ep_{i_1i_2} \equiv \ffaa
}
Diagram language is more comfortable and illustrative for higher writhe numbers, definitions of which we give below. 

It turns out that the second Vassiliev invariant cannot be expressed through the just defined writhe number. Hence, we modify the definition of the writhe number. First, we introduce two types of the crossing signs $\ep$.
\dfnl{eps}{
\eq{
\ep_{i_1i_2}^{ou} =
\left\{
\begin{array}{l}
  \ep_{i_1i_2} , \hspace{5mm}  if \ the \ strand, \ which \ led \ to \ the \ point \ (i_1i_2), \ is \ \textbf{over} \ another \ strand \\[4mm]
  0 , \hspace{1cm} otherwise \\[4mm]
\end{array}
\right.
}
\eq{
\ep_{i_1i_2}^{uo} =
\left\{
\begin{array}{l}
  \ep_{i_1i_2} , \hspace{5mm}  if \ the \ strand, \ which \ led \ to \ the \ point \ (i_1i_2), \ is \ \textbf{under} \ another \ strand \\[4mm]
  0 , \hspace{1cm} otherwise \\[4mm]
\end{array}
\right.
}
}
Thus based on the last definition we can define two types of colored writhe numbers.
\dfnl{wrt}{
Colored writhe numbers for oriented knot  are defined as the following sums
\eqs{
w_1^{ou} := \summ{i_1<i_2}{} \ep_{i_1i_2}^{ou}  \\
w_1^{uo} := \summ{i_1<i_2}{} \ep_{i_1i_2}^{uo}
}
}
Colored writhe numbers can be denoted by the following diagrams:
\vspace{4mm}
\eq{
w_1^{ou} \equiv \oau, \hspace{1cm}
w_1^{uo} \equiv \uao
}
\vspace{3mm}
\begin{remark}
 We always read chord diagrams clockwise from the top. It is equivalent to introducing an origin point $O$ slightly to the right from the top.
\end{remark}

\subsection{Higher writhe numbers}
Any definition from previous subSection can be generalized in a straightforward way. However because the colored writhe numbers are more general objects than the ordinary ones, we generalize only the last definition \re{wrt}. 

\dfnl{hw_n}{
Higher colored writhe number of $n$-th order for oriented knot $w(\K)$ is a sum
\eq{
w^{\sigma_1..\sigma_n}_n(\K) := \summ{i_1< \dots <i_{2n}}{} \ep_{i_{m_1}i_{m_2}}^{\sigma_1} \dots \ep_{i_{m_{2n-1}}i_{m_{2n}}}^{\sigma_n},
}
where $\sigma_1, \dots, \sigma_n$ take values $\rm{ou}$ or $\rm{uo}$.
}
In graphical approach indices $m_1,...,m_{2n}$ correspond to the points on the circle, $\ep_{i_{m_k}i_{m_l}}$ corresponds to a line between points $m_k$ and $m_l$ (see the following example).
\ex{
For order $2$ we have only two diagrams:
\eqs{
\summ{i_1<i_2<i_3<i_4}{} \ep_{i_1i_2} \ep_{i_3i_4} = \ffab \\
\summ{i_1<i_2<i_3<i_4}{} \ep_{i_1i_3} \ep_{i_2i_4} = \ffac
}
}

\subsection{Relations between higher writhe numbers}
Higher colored  writhe numbers are not independent. There are linear relations between them. These relations complicate the study of Vassiliev invariants because often it is difficult to reduce answers and understand their combinatorial context. For illustrative purpose we list here relations for orders $2$, $3$ and $4$.
\subsubsection{Order 2}
There is the only relation:
\vspace{7mm}
\eq{
\ouuo = \uoou
}
\subsubsection{Order 3}
There are already 22 relations:
\begin{align}
\; -\; \mathpic{0.4}{\draw[gray] (0,0) circle (1); \draw[->,red,thick] (-0.866025,0.5) -- (0.,1.); \draw[->,red,thick] (0.866025,0.5) -- (0.,-1.); \draw[->,red,thick] (-0.866025,-0.5) -- (0.866025,-0.5); }\; +\; \mathpic{0.4}{\draw[gray] (0,0) circle (1); \draw[->,red,thick] (-0.866025,0.5) -- (0.,1.); \draw[->,red,thick] (0.,-1.) -- (0.866025,0.5); \draw[->,red,thick] (0.866025,-0.5) -- (-0.866025,-0.5); }&=0 \\ 
\; -\; \mathpic{0.4}{\draw[gray] (0,0) circle (1); \draw[->,red,thick] (0.,1.) -- (-0.866025,0.5); \draw[->,red,thick] (0.866025,0.5) -- (0.,-1.); \draw[->,red,thick] (-0.866025,-0.5) -- (0.866025,-0.5); }\; +\; \mathpic{0.4}{\draw[gray] (0,0) circle (1); \draw[->,red,thick] (0.,1.) -- (-0.866025,0.5); \draw[->,red,thick] (0.,-1.) -- (0.866025,0.5); \draw[->,red,thick] (0.866025,-0.5) -- (-0.866025,-0.5); }&=0 \\ 
\; -\; \mathpic{0.4}{\draw[gray] (0,0) circle (1); \draw[->,red,thick] (0.866025,-0.5) -- (0.,1.); \draw[->,red,thick] (0.866025,0.5) -- (-0.866025,-0.5); \draw[->,red,thick] (-0.866025,0.5) -- (0.,-1.); }\; +\; \mathpic{0.4}{\draw[gray] (0,0) circle (1); \draw[->,red,thick] (0.866025,-0.5) -- (0.,1.); \draw[->,red,thick] (-0.866025,-0.5) -- (0.866025,0.5); \draw[->,red,thick] (-0.866025,0.5) -- (0.,-1.); }&=0 \\ 
\; -\; \mathpic{0.4}{\draw[gray] (0,0) circle (1); \draw[->,red,thick] (0.,1.) -- (0.866025,-0.5); \draw[->,red,thick] (0.866025,0.5) -- (-0.866025,-0.5); \draw[->,red,thick] (-0.866025,0.5) -- (0.,-1.); }\; +\; \mathpic{0.4}{\draw[gray] (0,0) circle (1); \draw[->,red,thick] (0.,1.) -- (0.866025,-0.5); \draw[->,red,thick] (-0.866025,-0.5) -- (0.866025,0.5); \draw[->,red,thick] (-0.866025,0.5) -- (0.,-1.); }&=0 \\ 
\mathpic{0.4}{\draw[gray] (0,0) circle (1); \draw[->,red,thick] (0.,1.) -- (0.,-1.); \draw[->,red,thick] (-0.866025,0.5) -- (0.866025,0.5); \draw[->,red,thick] (-0.866025,-0.5) -- (0.866025,-0.5); }\; -\; \mathpic{0.4}{\draw[gray] (0,0) circle (1); \draw[->,red,thick] (0.,-1.) -- (0.,1.); \draw[->,red,thick] (-0.866025,0.5) -- (0.866025,0.5); \draw[->,red,thick] (-0.866025,-0.5) -- (0.866025,-0.5); }\; -\; \mathpic{0.4}{\draw[gray] (0,0) circle (1); \draw[->,red,thick] (-0.866025,-0.5) -- (0.,1.); \draw[->,red,thick] (0.,-1.) -- (0.866025,0.5); \draw[->,red,thick] (0.866025,-0.5) -- (-0.866025,0.5); }\; +\; \mathpic{0.4}{\draw[gray] (0,0) circle (1); \draw[->,red,thick] (-0.866025,-0.5) -- (0.,1.); \draw[->,red,thick] (0.,-1.) -- (0.866025,0.5); \draw[->,red,thick] (-0.866025,0.5) -- (0.866025,-0.5); }&=0 \\ 
\mathpic{0.4}{\draw[gray] (0,0) circle (1); \draw[->,red,thick] (0.,1.) -- (0.,-1.); \draw[->,red,thick] (-0.866025,0.5) -- (0.866025,0.5); \draw[->,red,thick] (0.866025,-0.5) -- (-0.866025,-0.5); }\; -\; \mathpic{0.4}{\draw[gray] (0,0) circle (1); \draw[->,red,thick] (0.,-1.) -- (0.,1.); \draw[->,red,thick] (-0.866025,0.5) -- (0.866025,0.5); \draw[->,red,thick] (0.866025,-0.5) -- (-0.866025,-0.5); }\; -\; \mathpic{0.4}{\draw[gray] (0,0) circle (1); \draw[->,red,thick] (-0.866025,-0.5) -- (0.,1.); \draw[->,red,thick] (0.866025,0.5) -- (0.,-1.); \draw[->,red,thick] (0.866025,-0.5) -- (-0.866025,0.5); }\; +\; \mathpic{0.4}{\draw[gray] (0,0) circle (1); \draw[->,red,thick] (-0.866025,-0.5) -- (0.,1.); \draw[->,red,thick] (0.866025,0.5) -- (0.,-1.); \draw[->,red,thick] (-0.866025,0.5) -- (0.866025,-0.5); }&=0 \\ 
\mathpic{0.4}{\draw[gray] (0,0) circle (1); \draw[->,red,thick] (0.,1.) -- (0.,-1.); \draw[->,red,thick] (0.866025,0.5) -- (-0.866025,0.5); \draw[->,red,thick] (-0.866025,-0.5) -- (0.866025,-0.5); }\; -\; \mathpic{0.4}{\draw[gray] (0,0) circle (1); \draw[->,red,thick] (0.,-1.) -- (0.,1.); \draw[->,red,thick] (0.866025,0.5) -- (-0.866025,0.5); \draw[->,red,thick] (-0.866025,-0.5) -- (0.866025,-0.5); }\; -\; \mathpic{0.4}{\draw[gray] (0,0) circle (1); \draw[->,red,thick] (0.,1.) -- (-0.866025,-0.5); \draw[->,red,thick] (0.,-1.) -- (0.866025,0.5); \draw[->,red,thick] (0.866025,-0.5) -- (-0.866025,0.5); }\; +\; \mathpic{0.4}{\draw[gray] (0,0) circle (1); \draw[->,red,thick] (0.,1.) -- (-0.866025,-0.5); \draw[->,red,thick] (0.,-1.) -- (0.866025,0.5); \draw[->,red,thick] (-0.866025,0.5) -- (0.866025,-0.5); }&=0 \\ 
\; -\; \mathpic{0.4}{\draw[gray] (0,0) circle (1); \draw[->,red,thick] (0.,1.) -- (0.,-1.); \draw[->,red,thick] (0.866025,0.5) -- (-0.866025,0.5); \draw[->,red,thick] (0.866025,-0.5) -- (-0.866025,-0.5); }\; +\; \mathpic{0.4}{\draw[gray] (0,0) circle (1); \draw[->,red,thick] (0.,-1.) -- (0.,1.); \draw[->,red,thick] (0.866025,0.5) -- (-0.866025,0.5); \draw[->,red,thick] (0.866025,-0.5) -- (-0.866025,-0.5); }\; +\; \mathpic{0.4}{\draw[gray] (0,0) circle (1); \draw[->,red,thick] (0.,1.) -- (-0.866025,-0.5); \draw[->,red,thick] (0.866025,0.5) -- (0.,-1.); \draw[->,red,thick] (0.866025,-0.5) -- (-0.866025,0.5); }\; -\; \mathpic{0.4}{\draw[gray] (0,0) circle (1); \draw[->,red,thick] (0.,1.) -- (-0.866025,-0.5); \draw[->,red,thick] (0.866025,0.5) -- (0.,-1.); \draw[->,red,thick] (-0.866025,0.5) -- (0.866025,-0.5); }&=0 \\ 
\mathpic{0.4}{\draw[gray] (0,0) circle (1); \draw[->,red,thick] (0.,1.) -- (-0.866025,-0.5); \draw[->,red,thick] (0.866025,0.5) -- (0.866025,-0.5); \draw[->,red,thick] (0.,-1.) -- (-0.866025,0.5); }\; -\; \mathpic{0.4}{\draw[gray] (0,0) circle (1); \draw[->,red,thick] (0.,1.) -- (-0.866025,-0.5); \draw[->,red,thick] (0.866025,0.5) -- (0.866025,-0.5); \draw[->,red,thick] (-0.866025,0.5) -- (0.,-1.); }\; -\; \mathpic{0.4}{\draw[gray] (0,0) circle (1); \draw[->,red,thick] (-0.866025,-0.5) -- (0.,1.); \draw[->,red,thick] (0.866025,0.5) -- (0.866025,-0.5); \draw[->,red,thick] (0.,-1.) -- (-0.866025,0.5); }\; +\; \mathpic{0.4}{\draw[gray] (0,0) circle (1); \draw[->,red,thick] (-0.866025,-0.5) -- (0.,1.); \draw[->,red,thick] (0.866025,0.5) -- (0.866025,-0.5); \draw[->,red,thick] (-0.866025,0.5) -- (0.,-1.); }&=0 \\ 
\mathpic{0.4}{\draw[gray] (0,0) circle (1); \draw[->,red,thick] (0.,1.) -- (0.866025,-0.5); \draw[->,red,thick] (0.866025,0.5) -- (-0.866025,0.5); \draw[->,red,thick] (-0.866025,-0.5) -- (0.,-1.); }\; -\; \mathpic{0.4}{\draw[gray] (0,0) circle (1); \draw[->,red,thick] (0.,1.) -- (0.866025,-0.5); \draw[->,red,thick] (-0.866025,0.5) -- (0.866025,0.5); \draw[->,red,thick] (-0.866025,-0.5) -- (0.,-1.); }\; -\; \mathpic{0.4}{\draw[gray] (0,0) circle (1); \draw[->,red,thick] (0.866025,-0.5) -- (0.,1.); \draw[->,red,thick] (0.866025,0.5) -- (-0.866025,0.5); \draw[->,red,thick] (-0.866025,-0.5) -- (0.,-1.); }\; +\; \mathpic{0.4}{\draw[gray] (0,0) circle (1); \draw[->,red,thick] (0.866025,-0.5) -- (0.,1.); \draw[->,red,thick] (-0.866025,0.5) -- (0.866025,0.5); \draw[->,red,thick] (-0.866025,-0.5) -- (0.,-1.); }&=0 \\ 
\mathpic{0.4}{\draw[gray] (0,0) circle (1); \draw[->,red,thick] (0.,1.) -- (0.866025,-0.5); \draw[->,red,thick] (0.866025,0.5) -- (-0.866025,0.5); \draw[->,red,thick] (0.,-1.) -- (-0.866025,-0.5); }\; -\; \mathpic{0.4}{\draw[gray] (0,0) circle (1); \draw[->,red,thick] (0.,1.) -- (0.866025,-0.5); \draw[->,red,thick] (-0.866025,0.5) -- (0.866025,0.5); \draw[->,red,thick] (0.,-1.) -- (-0.866025,-0.5); }\; -\; \mathpic{0.4}{\draw[gray] (0,0) circle (1); \draw[->,red,thick] (0.866025,-0.5) -- (0.,1.); \draw[->,red,thick] (0.866025,0.5) -- (-0.866025,0.5); \draw[->,red,thick] (0.,-1.) -- (-0.866025,-0.5); }\; +\; \mathpic{0.4}{\draw[gray] (0,0) circle (1); \draw[->,red,thick] (0.866025,-0.5) -- (0.,1.); \draw[->,red,thick] (-0.866025,0.5) -- (0.866025,0.5); \draw[->,red,thick] (0.,-1.) -- (-0.866025,-0.5); }&=0 \\ 
\; -\; \mathpic{0.4}{\draw[gray] (0,0) circle (1); \draw[->,red,thick] (0.,1.) -- (0.866025,-0.5); \draw[->,red,thick] (0.,-1.) -- (0.866025,0.5); \draw[->,red,thick] (-0.866025,0.5) -- (-0.866025,-0.5); }\; +\; \mathpic{0.4}{\draw[gray] (0,0) circle (1); \draw[->,red,thick] (0.866025,-0.5) -- (0.,1.); \draw[->,red,thick] (0.866025,0.5) -- (0.,-1.); \draw[->,red,thick] (-0.866025,0.5) -- (-0.866025,-0.5); }\; +\; \mathpic{0.4}{\draw[gray] (0,0) circle (1); \draw[->,red,thick] (0.,1.) -- (0.866025,-0.5); \draw[->,red,thick] (0.866025,0.5) -- (-0.866025,0.5); \draw[->,red,thick] (-0.866025,-0.5) -- (0.,-1.); }\; -\; \mathpic{0.4}{\draw[gray] (0,0) circle (1); \draw[->,red,thick] (0.,1.) -- (0.866025,-0.5); \draw[->,red,thick] (-0.866025,0.5) -- (0.866025,0.5); \draw[->,red,thick] (-0.866025,-0.5) -- (0.,-1.); }&=0 \\ 
\; -\; \mathpic{0.4}{\draw[gray] (0,0) circle (1); \draw[->,red,thick] (0.,1.) -- (0.866025,-0.5); \draw[->,red,thick] (0.,-1.) -- (0.866025,0.5); \draw[->,red,thick] (-0.866025,-0.5) -- (-0.866025,0.5); }\; +\; \mathpic{0.4}{\draw[gray] (0,0) circle (1); \draw[->,red,thick] (0.866025,-0.5) -- (0.,1.); \draw[->,red,thick] (0.866025,0.5) -- (0.,-1.); \draw[->,red,thick] (-0.866025,-0.5) -- (-0.866025,0.5); }\; +\; \mathpic{0.4}{\draw[gray] (0,0) circle (1); \draw[->,red,thick] (0.866025,-0.5) -- (0.,1.); \draw[->,red,thick] (0.866025,0.5) -- (-0.866025,0.5); \draw[->,red,thick] (0.,-1.) -- (-0.866025,-0.5); }\; -\; \mathpic{0.4}{\draw[gray] (0,0) circle (1); \draw[->,red,thick] (0.866025,-0.5) -- (0.,1.); \draw[->,red,thick] (-0.866025,0.5) -- (0.866025,0.5); \draw[->,red,thick] (0.,-1.) -- (-0.866025,-0.5); }&=0 \\ 
\; -\; \mathpic{0.4}{\draw[gray] (0,0) circle (1); \draw[->,red,thick] (0.866025,0.5) -- (0.,1.); \draw[->,red,thick] (0.866025,-0.5) -- (-0.866025,-0.5); \draw[->,red,thick] (-0.866025,0.5) -- (0.,-1.); }\; +\; \mathpic{0.4}{\draw[gray] (0,0) circle (1); \draw[->,red,thick] (0.866025,0.5) -- (0.,1.); \draw[->,red,thick] (-0.866025,-0.5) -- (0.866025,-0.5); \draw[->,red,thick] (0.,-1.) -- (-0.866025,0.5); }\; -\; \mathpic{0.4}{\draw[gray] (0,0) circle (1); \draw[->,red,thick] (0.,1.) -- (-0.866025,-0.5); \draw[->,red,thick] (0.866025,-0.5) -- (0.866025,0.5); \draw[->,red,thick] (0.,-1.) -- (-0.866025,0.5); }\; +\; \mathpic{0.4}{\draw[gray] (0,0) circle (1); \draw[->,red,thick] (-0.866025,-0.5) -- (0.,1.); \draw[->,red,thick] (0.866025,-0.5) -- (0.866025,0.5); \draw[->,red,thick] (0.,-1.) -- (-0.866025,0.5); }&=0 \\ 
\mathpic{0.4}{\draw[gray] (0,0) circle (1); \draw[->,red,thick] (0.,1.) -- (0.866025,0.5); \draw[->,red,thick] (0.866025,-0.5) -- (-0.866025,-0.5); \draw[->,red,thick] (-0.866025,0.5) -- (0.,-1.); }\; -\; \mathpic{0.4}{\draw[gray] (0,0) circle (1); \draw[->,red,thick] (0.,1.) -- (0.866025,0.5); \draw[->,red,thick] (-0.866025,-0.5) -- (0.866025,-0.5); \draw[->,red,thick] (0.,-1.) -- (-0.866025,0.5); }\; +\; \mathpic{0.4}{\draw[gray] (0,0) circle (1); \draw[->,red,thick] (0.,1.) -- (-0.866025,-0.5); \draw[->,red,thick] (0.866025,0.5) -- (0.866025,-0.5); \draw[->,red,thick] (-0.866025,0.5) -- (0.,-1.); }\; -\; \mathpic{0.4}{\draw[gray] (0,0) circle (1); \draw[->,red,thick] (-0.866025,-0.5) -- (0.,1.); \draw[->,red,thick] (0.866025,0.5) -- (0.866025,-0.5); \draw[->,red,thick] (-0.866025,0.5) -- (0.,-1.); }&=0 \\ 
\; -\; \mathpic{0.4}{\draw[gray] (0,0) circle (1); \draw[->,red,thick] (0.,1.) -- (0.866025,-0.5); \draw[->,red,thick] (-0.866025,0.5) -- (0.866025,0.5); \draw[->,red,thick] (0.,-1.) -- (-0.866025,-0.5); }\; +\; \mathpic{0.4}{\draw[gray] (0,0) circle (1); \draw[->,red,thick] (0.866025,-0.5) -- (0.,1.); \draw[->,red,thick] (-0.866025,0.5) -- (0.866025,0.5); \draw[->,red,thick] (0.,-1.) -- (-0.866025,-0.5); }\; +\; \mathpic{0.4}{\draw[gray] (0,0) circle (1); \draw[->,red,thick] (-0.866025,-0.5) -- (0.,1.); \draw[->,red,thick] (0.866025,0.5) -- (0.866025,-0.5); \draw[->,red,thick] (0.,-1.) -- (-0.866025,0.5); }\; -\; \mathpic{0.4}{\draw[gray] (0,0) circle (1); \draw[->,red,thick] (-0.866025,-0.5) -- (0.,1.); \draw[->,red,thick] (0.866025,0.5) -- (0.866025,-0.5); \draw[->,red,thick] (-0.866025,0.5) -- (0.,-1.); }\; -\; \mathpic{0.4}{\draw[gray] (0,0) circle (1); \draw[->,red,thick] (0.,1.) -- (-0.866025,-0.5); \draw[->,red,thick] (-0.866025,0.5) -- (0.866025,0.5); \draw[->,red,thick] (0.866025,-0.5) -- (0.,-1.); }\; +\; \mathpic{0.4}{\draw[gray] (0,0) circle (1); \draw[->,red,thick] (-0.866025,-0.5) -- (0.,1.); \draw[->,red,thick] (0.866025,0.5) -- (-0.866025,0.5); \draw[->,red,thick] (0.866025,-0.5) -- (0.,-1.); }&=0 \\ 
\mathpic{0.4}{\draw[gray] (0,0) circle (1); \draw[->,red,thick] (0.,1.) -- (0.866025,-0.5); \draw[->,red,thick] (-0.866025,-0.5) -- (0.866025,0.5); \draw[->,red,thick] (0.,-1.) -- (-0.866025,0.5); }\; -\; \mathpic{0.4}{\draw[gray] (0,0) circle (1); \draw[->,red,thick] (0.866025,-0.5) -- (0.,1.); \draw[->,red,thick] (0.866025,0.5) -- (-0.866025,-0.5); \draw[->,red,thick] (-0.866025,0.5) -- (0.,-1.); }\; -\; \mathpic{0.4}{\draw[gray] (0,0) circle (1); \draw[->,red,thick] (0.,-1.) -- (0.,1.); \draw[->,red,thick] (0.866025,0.5) -- (-0.866025,0.5); \draw[->,red,thick] (-0.866025,-0.5) -- (0.866025,-0.5); }\; +\; \mathpic{0.4}{\draw[gray] (0,0) circle (1); \draw[->,red,thick] (0.,-1.) -- (0.,1.); \draw[->,red,thick] (-0.866025,0.5) -- (0.866025,0.5); \draw[->,red,thick] (0.866025,-0.5) -- (-0.866025,-0.5); }\; -\; \mathpic{0.4}{\draw[gray] (0,0) circle (1); \draw[->,red,thick] (0.,1.) -- (-0.866025,-0.5); \draw[->,red,thick] (0.,-1.) -- (0.866025,0.5); \draw[->,red,thick] (0.866025,-0.5) -- (-0.866025,0.5); }\; +\; \mathpic{0.4}{\draw[gray] (0,0) circle (1); \draw[->,red,thick] (-0.866025,-0.5) -- (0.,1.); \draw[->,red,thick] (0.866025,0.5) -- (0.,-1.); \draw[->,red,thick] (0.866025,-0.5) -- (-0.866025,0.5); }&=0 \\ 
\mathpic{0.4}{\draw[gray] (0,0) circle (1); \draw[->,red,thick] (0.,1.) -- (0.866025,-0.5); \draw[->,red,thick] (0.866025,0.5) -- (-0.866025,-0.5); \draw[->,red,thick] (0.,-1.) -- (-0.866025,0.5); }\; -\; \mathpic{0.4}{\draw[gray] (0,0) circle (1); \draw[->,red,thick] (0.866025,-0.5) -- (0.,1.); \draw[->,red,thick] (0.866025,0.5) -- (-0.866025,-0.5); \draw[->,red,thick] (-0.866025,0.5) -- (0.,-1.); }\; -\; \mathpic{0.4}{\draw[gray] (0,0) circle (1); \draw[->,red,thick] (0.,-1.) -- (0.,1.); \draw[->,red,thick] (0.866025,0.5) -- (-0.866025,0.5); \draw[->,red,thick] (-0.866025,-0.5) -- (0.866025,-0.5); }\; +\; \mathpic{0.4}{\draw[gray] (0,0) circle (1); \draw[->,red,thick] (0.,-1.) -- (0.,1.); \draw[->,red,thick] (-0.866025,0.5) -- (0.866025,0.5); \draw[->,red,thick] (0.866025,-0.5) -- (-0.866025,-0.5); }\; -\; \mathpic{0.4}{\draw[gray] (0,0) circle (1); \draw[->,red,thick] (0.,1.) -- (-0.866025,-0.5); \draw[->,red,thick] (0.,-1.) -- (0.866025,0.5); \draw[->,red,thick] (0.866025,-0.5) -- (-0.866025,0.5); }\; +\; \mathpic{0.4}{\draw[gray] (0,0) circle (1); \draw[->,red,thick] (-0.866025,-0.5) -- (0.,1.); \draw[->,red,thick] (0.866025,0.5) -- (0.,-1.); \draw[->,red,thick] (0.866025,-0.5) -- (-0.866025,0.5); }&=0 \\ 
\mathpic{0.4}{\draw[gray] (0,0) circle (1); \draw[->,red,thick] (0.,1.) -- (0.866025,-0.5); \draw[->,red,thick] (0.866025,0.5) -- (-0.866025,0.5); \draw[->,red,thick] (-0.866025,-0.5) -- (0.,-1.); }\; -\; \mathpic{0.4}{\draw[gray] (0,0) circle (1); \draw[->,red,thick] (0.866025,-0.5) -- (0.,1.); \draw[->,red,thick] (0.866025,0.5) -- (-0.866025,0.5); \draw[->,red,thick] (-0.866025,-0.5) -- (0.,-1.); }\; -\; \mathpic{0.4}{\draw[gray] (0,0) circle (1); \draw[->,red,thick] (-0.866025,-0.5) -- (0.,1.); \draw[->,red,thick] (0.866025,-0.5) -- (0.866025,0.5); \draw[->,red,thick] (0.,-1.) -- (-0.866025,0.5); }\; +\; \mathpic{0.4}{\draw[gray] (0,0) circle (1); \draw[->,red,thick] (-0.866025,-0.5) -- (0.,1.); \draw[->,red,thick] (0.866025,-0.5) -- (0.866025,0.5); \draw[->,red,thick] (-0.866025,0.5) -- (0.,-1.); }\; +\; \mathpic{0.4}{\draw[gray] (0,0) circle (1); \draw[->,red,thick] (0.,1.) -- (-0.866025,-0.5); \draw[->,red,thick] (-0.866025,0.5) -- (0.866025,0.5); \draw[->,red,thick] (0.,-1.) -- (0.866025,-0.5); }\; -\; \mathpic{0.4}{\draw[gray] (0,0) circle (1); \draw[->,red,thick] (-0.866025,-0.5) -- (0.,1.); \draw[->,red,thick] (0.866025,0.5) -- (-0.866025,0.5); \draw[->,red,thick] (0.,-1.) -- (0.866025,-0.5); }&=0 \\ 
\mathpic{0.4}{\draw[gray] (0,0) circle (1); \draw[->,red,thick] (0.,1.) -- (0.866025,-0.5); \draw[->,red,thick] (0.866025,0.5) -- (-0.866025,0.5); \draw[->,red,thick] (-0.866025,-0.5) -- (0.,-1.); }\; -\; \mathpic{0.4}{\draw[gray] (0,0) circle (1); \draw[->,red,thick] (0.866025,-0.5) -- (0.,1.); \draw[->,red,thick] (0.866025,0.5) -- (-0.866025,0.5); \draw[->,red,thick] (-0.866025,-0.5) -- (0.,-1.); }\; -\; \mathpic{0.4}{\draw[gray] (0,0) circle (1); \draw[->,red,thick] (0.,1.) -- (-0.866025,-0.5); \draw[->,red,thick] (0.866025,-0.5) -- (0.866025,0.5); \draw[->,red,thick] (0.,-1.) -- (-0.866025,0.5); }\; +\; \mathpic{0.4}{\draw[gray] (0,0) circle (1); \draw[->,red,thick] (0.,1.) -- (-0.866025,-0.5); \draw[->,red,thick] (0.866025,-0.5) -- (0.866025,0.5); \draw[->,red,thick] (-0.866025,0.5) -- (0.,-1.); }\; +\; \mathpic{0.4}{\draw[gray] (0,0) circle (1); \draw[->,red,thick] (0.,1.) -- (-0.866025,-0.5); \draw[->,red,thick] (-0.866025,0.5) -- (0.866025,0.5); \draw[->,red,thick] (0.,-1.) -- (0.866025,-0.5); }\; -\; \mathpic{0.4}{\draw[gray] (0,0) circle (1); \draw[->,red,thick] (-0.866025,-0.5) -- (0.,1.); \draw[->,red,thick] (0.866025,0.5) -- (-0.866025,0.5); \draw[->,red,thick] (0.,-1.) -- (0.866025,-0.5); }&=0 \\ 
\mathpic{0.4}{\draw[gray] (0,0) circle (1); \draw[->,red,thick] (0.,1.) -- (0.866025,-0.5); \draw[->,red,thick] (0.866025,0.5) -- (-0.866025,-0.5); \draw[->,red,thick] (-0.866025,0.5) -- (0.,-1.); }\; -\; 2\mathpic{0.4}{\draw[gray] (0,0) circle (1); \draw[->,red,thick] (0.866025,-0.5) -- (0.,1.); \draw[->,red,thick] (0.866025,0.5) -- (-0.866025,-0.5); \draw[->,red,thick] (-0.866025,0.5) -- (0.,-1.); }\; +\; \mathpic{0.4}{\draw[gray] (0,0) circle (1); \draw[->,red,thick] (0.866025,-0.5) -- (0.,1.); \draw[->,red,thick] (-0.866025,-0.5) -- (0.866025,0.5); \draw[->,red,thick] (0.,-1.) -- (-0.866025,0.5); }\; +\; \mathpic{0.4}{\draw[gray] (0,0) circle (1); \draw[->,red,thick] (0.,1.) -- (0.,-1.); \draw[->,red,thick] (0.866025,0.5) -- (-0.866025,0.5); \draw[->,red,thick] (-0.866025,-0.5) -- (0.866025,-0.5); }\; -\; \mathpic{0.4}{\draw[gray] (0,0) circle (1); \draw[->,red,thick] (0.,1.) -- (0.,-1.); \draw[->,red,thick] (-0.866025,0.5) -- (0.866025,0.5); \draw[->,red,thick] (0.866025,-0.5) -- (-0.866025,-0.5); }\; -\; 2\mathpic{0.4}{\draw[gray] (0,0) circle (1); \draw[->,red,thick] (0.,-1.) -- (0.,1.); \draw[->,red,thick] (0.866025,0.5) -- (-0.866025,0.5); \draw[->,red,thick] (-0.866025,-0.5) -- (0.866025,-0.5); }\; +\; 2\mathpic{0.4}{\draw[gray] (0,0) circle (1); \draw[->,red,thick] (0.,-1.) -- (0.,1.); \draw[->,red,thick] (-0.866025,0.5) -- (0.866025,0.5); \draw[->,red,thick] (0.866025,-0.5) -- (-0.866025,-0.5); }\; -\; \mathpic{0.4}{\draw[gray] (0,0) circle (1); \draw[->,red,thick] (0.,1.) -- (-0.866025,-0.5); \draw[->,red,thick] (0.,-1.) -- (0.866025,0.5); \draw[->,red,thick] (0.866025,-0.5) -- (-0.866025,0.5); }\; +\; \mathpic{0.4}{\draw[gray] (0,0) circle (1); \draw[->,red,thick] (-0.866025,-0.5) -- (0.,1.); \draw[->,red,thick] (0.866025,0.5) -- (0.,-1.); \draw[->,red,thick] (0.866025,-0.5) -- (-0.866025,0.5); }&=0 \\ 
\mathpic{0.4}{\draw[gray] (0,0) circle (1); \draw[->,red,thick] (0.,1.) -- (0.866025,-0.5); \draw[->,red,thick] (0.866025,0.5) -- (-0.866025,-0.5); \draw[->,red,thick] (-0.866025,0.5) -- (0.,-1.); }\; +\; \mathpic{0.4}{\draw[gray] (0,0) circle (1); \draw[->,red,thick] (0.866025,-0.5) -- (0.,1.); \draw[->,red,thick] (0.866025,0.5) -- (-0.866025,-0.5); \draw[->,red,thick] (0.,-1.) -- (-0.866025,0.5); }\; -\; 2\mathpic{0.4}{\draw[gray] (0,0) circle (1); \draw[->,red,thick] (0.866025,-0.5) -- (0.,1.); \draw[->,red,thick] (0.866025,0.5) -- (-0.866025,-0.5); \draw[->,red,thick] (-0.866025,0.5) -- (0.,-1.); }\; +\; \mathpic{0.4}{\draw[gray] (0,0) circle (1); \draw[->,red,thick] (0.,1.) -- (0.,-1.); \draw[->,red,thick] (0.866025,0.5) -- (-0.866025,0.5); \draw[->,red,thick] (-0.866025,-0.5) -- (0.866025,-0.5); }\; -\; \mathpic{0.4}{\draw[gray] (0,0) circle (1); \draw[->,red,thick] (0.,1.) -- (0.,-1.); \draw[->,red,thick] (-0.866025,0.5) -- (0.866025,0.5); \draw[->,red,thick] (0.866025,-0.5) -- (-0.866025,-0.5); }\; -\; 2\mathpic{0.4}{\draw[gray] (0,0) circle (1); \draw[->,red,thick] (0.,-1.) -- (0.,1.); \draw[->,red,thick] (0.866025,0.5) -- (-0.866025,0.5); \draw[->,red,thick] (-0.866025,-0.5) -- (0.866025,-0.5); }\; +\; 2\mathpic{0.4}{\draw[gray] (0,0) circle (1); \draw[->,red,thick] (0.,-1.) -- (0.,1.); \draw[->,red,thick] (-0.866025,0.5) -- (0.866025,0.5); \draw[->,red,thick] (0.866025,-0.5) -- (-0.866025,-0.5); }\; -\; \mathpic{0.4}{\draw[gray] (0,0) circle (1); \draw[->,red,thick] (0.,1.) -- (-0.866025,-0.5); \draw[->,red,thick] (0.,-1.) -- (0.866025,0.5); \draw[->,red,thick] (0.866025,-0.5) -- (-0.866025,0.5); }\; +\; \mathpic{0.4}{\draw[gray] (0,0) circle (1); \draw[->,red,thick] (-0.866025,-0.5) -- (0.,1.); \draw[->,red,thick] (0.866025,0.5) -- (0.,-1.); \draw[->,red,thick] (0.866025,-0.5) -- (-0.866025,0.5); }&=0 
\end{align}

Let us note that all of the relations do not mix diagrams with different topology.
\subsubsection{Order 4}
There are 195 relations and we do not list all of them. We just note that in order 4 relations do not mix connected and disconnected diagrams and diagrams with different topology.

\subsection{Vassiliev invariants via higher writhe numbers}
With the help of higher colored writhe numbers one can give very easy and elegant combinatorial formulas for Vassiliev invariants. 

\subsubsection{Numerical experimental data}
From our numerical results (mentioned in Section \ref{NumRes}) we find some useful properties of Vassiliev invariants. All of the results in this subSection are purely experimental ones.
\begin{itemize}
\item Vassiliev invariants are can always be represented by sums of colored  writhes with symmetric chord diagrams with respect to change of all over-crossings to under-crossings and under-crossings to over-crossings. It follows from the fact that Vassiliev invariants do not change under reflection. We call such combinations \textit{reflective}.  Therefore it is convenient to introduce reflective combinations of writhes as follows:

\vspace{8mm}
$$
\oap =\oau +\uao $$ \\ $$
\hskip10pt \tapp = \ouuo + \uoou 
$$
\vspace{8mm}

\item Vassiliev invariants can always be represented by sums of writhes with connected chord diagrams. For example, the first diagram in
the figure below is connected while the second is not:

 \vspace{8mm}
$$
\tapp,\ \ \ \tbpp
$$
 \vspace{
8mm}

 \item Vassiliev invariants are
always equal to a sum of writhes whose chord diagrams are
\textit{irreducible}, which means that we can not stick together any
two chords in the diagram. Here we implicitly assume that to the left of the top point of the diagram we have the basepoint. Chords separated by the basepoint cannot be stuck together. For example, the first diagram in the figure
below is reducible and the second one is irreducible:

 \vspace{
8mm}
$$
\tappp,\ \ \ \tappm
$$
\vspace{ 8mm}

This property actually means that Vassiliev invariants of order $k$ do not mix with invariants of order less than $k$.
\end{itemize}

\subsubsection{Order 2}
One confusing fact related to writhes is their dependence on the choice of
the base point on the knot projection. To construct invariants we
should use only invariant combination of writhes that do not
depend on such a choice. Invariant combination of writhes with two
chords are the following ones:
\bigskip
\bigskip
\bigskip
$$
C_{1}=\tapp\ \ \ \ C_{2}=\tapm
$$
\bigskip
\bigskip
\bigskip
$$
C_{3}=\tbpp+\tcpm\ \ \ C_{4}=\tcpp+\tbpm
$$

\vspace{ 7mm}

The second Vassiliev invariant should be some linear combination
of these four expressions. We find:
\bigskip
\bigskip
\eql{vas2}{
{\cal{V}}_{2,1}=\tapm
}
Note that due to the presence of the basepoint this diagram is
irreducible.

\subsubsection{Order 3} 
Invariant combinations of irreducible and connected writhes are the following ones: \vspace{
7mm}
$$
C_{1}=\tcppp+\tapmp+\tbpmp
$$
\vspace{ 11mm}
$$
C_{2}=\tcppm+\tappm+\tbpmp
$$
\vspace{ 11mm}
$$
C_{3}=\tcpmp+\tappm+\tbpmp
$$
\vspace{ 11mm}
$$
C_{4}=\tcpmp+\tapmp+\tbpmm
$$
\vspace{ 11mm}
$$
C_{5}=\tcppm+\tapmp+\tbpmm
$$
\vspace{ 11mm}
$$
C_{6}=\tdpmp
$$
\vspace{ 3mm}

 We found the following interesting property of the
first five combinations: \be
\label{id1}C_{1}=C_{2}=C_{3}=C_{4}=C_{5} \ee

For the third Vassiliev invariant we get:
\eql{vas3}{
{\cal{V}}_{3,1}=\dfrac{1}{2}\,C_{1}+C_{6}
}

\section*{Acknowledgements}

We are grateful to A.Anokhina, E.Krylov, I.Polyubin, A.Popolitov, V.Slepukhin, Ye.Zenkevich and all participants of D.Anosov seminar for helpful remarks. We are also grateful to D.Bar-Natan for publishing part of our numerical results on the site katlas.org. We are especially grateful to A.Morozov for stimulating discussions.

Our work is partly supported by Ministry of Education and Science of the Russian Federation under contracts
14.740.11.0081 (P.DB., A.Sl.) and 14.740.11.0347 (A.Sm.), by RFBR grants 10-02-00499 (P.DB.), 10-01-00536 (A.Sl., A.Sm.) and 09-01-93106-NCNILa (A.Sm.), by joint grants 09-02-91005-ANF (P.DB., A.Sl.), 10-02-92109-Yaf-a (P.DB.), 11-01-92612-Royal Society(P.DB., A.Sl., A.Sm.), by NWO grant 613.001.021 (P.DB.) and by Dynasty Foundation.

\appendix

\section{KI combinatorially for figure-eight knot}
\label{app:fek}

In this appendix we will discuss the combinatorial technique of computing KI coefficients in more detail for the case of the figure-eight knot.

\begin{figure}[h!]
  \vspace{-15pt}
  \begin{center}
    \includegraphics[scale=0.71]{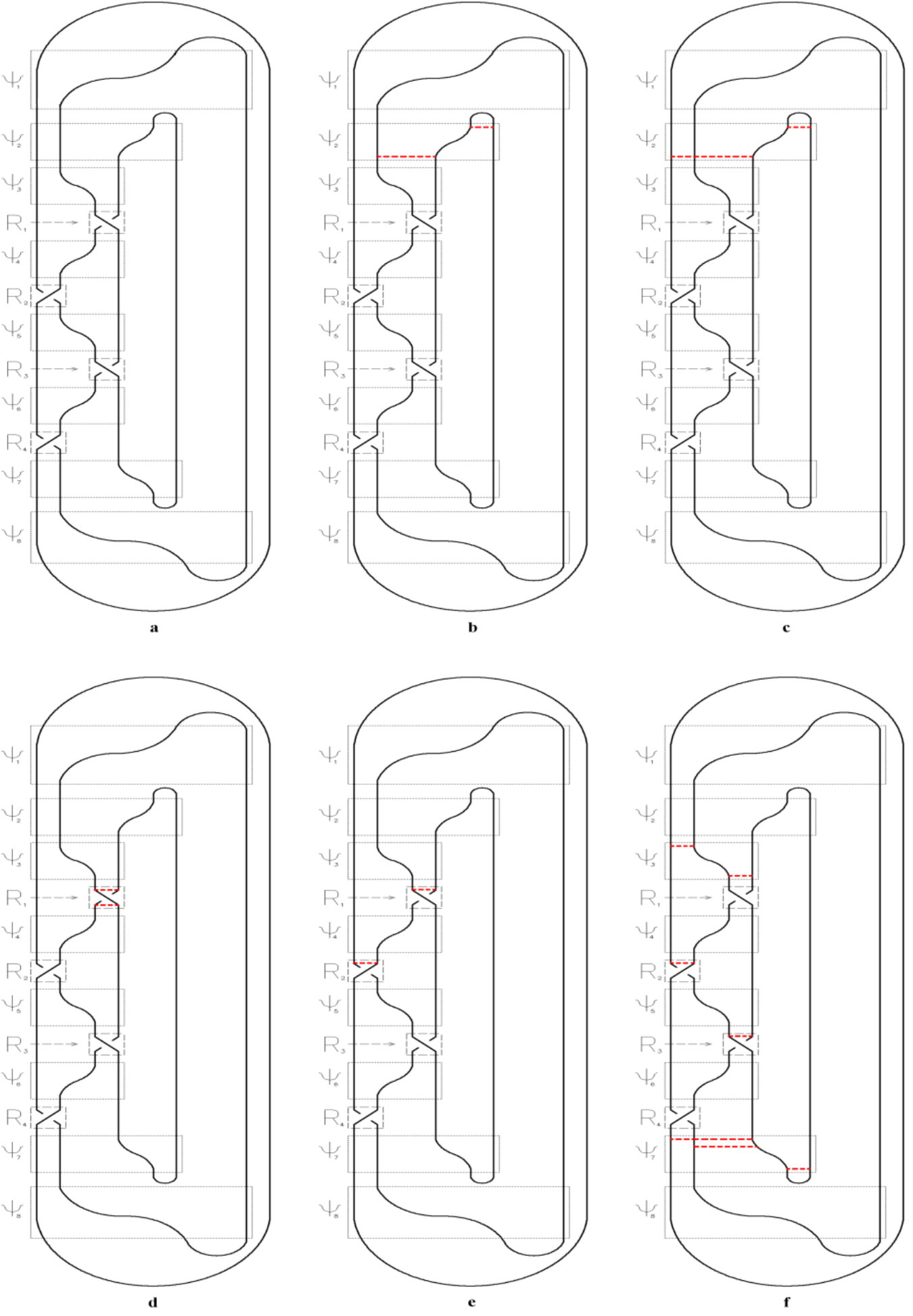}
  \end{center}
  \vspace{-15pt}
  \caption{Propagators}
\label{fig:prop}
\end{figure}

Let $K$ be the figure-eight knot, i.e. knot $4_1$.
Braid representation for $K$ with associators in correct positions is given in Figure \ref{fig:prop}a.
There are 8 associators $\Psi_1,\dots,\Psi_8$ and 4 R-matrices $R_1,\dots,R_4$. Corresponding Sections of the knot are enclosed in thin dotted and dashed boxes. 

Let us describe the computation of KI coefficients of order 2.

Recall that first of all one has to choose an integer partition of $n=2$ into 12 parts (corresponding to R-matrices and associators). Note that there is no linear in $T^aT^a$ term in \re{assoc}, so all partitions $n=\phi_1+\dots+\phi_8+r_1+\dots+r_4$ with any of $\phi_i$ equal to 1 give vanishing contributions.

Then we will have 4 partitions where both chords are taken from one of the R-matrices. One of the cases is drawn in Figure \ref{fig:prop}d, where the propagators, i.e. chords, are represented by red dashed lines (the thicker ones). Further on, we will have 6 partitions where the chords are taken from two different R-matrices (Figure \ref{fig:prop}e) and 8 partitions where both chords are taken from one of the associators. This latter case is more complicated than the former ones, where we had only one term of a given order, according to \re{rmat}. In the case of associators, however, we will have two complications: first, formula \re{assoc} has already several terms of a given order and, second, the number of terms is further multiplied in accordance with (\ref{reqpr}) if we have lines to the left, as in the cases of $\Psi_2$ and $\Psi_7$. Actually, for the order 2 we have just two relevant terms in \re{assoc}, the one drawn in Figure \ref{fig:prop}b and the one which differs from it by interchanging their endpoints on the central string. Formula (\ref{reqpr}) implies that one should also include the same terms but with the leftmost endpoint shifted to the leftmost string, like in Figure \ref{fig:prop}c.

The terms in KI corresponding to the choice of chords as in Figure \ref{fig:prop} are as follows:
\begin{enumerate}
 \item Figure \ref{fig:prop}b: $\displaystyle \dfrac{1}{24}\br{T^aT^aT^bT^b}$,
\item Figure \ref{fig:prop}c: $\displaystyle \dfrac{1}{24}\br{T^aT^aT^bT^b}$,
\item Figure \ref{fig:prop}d: $\displaystyle \dfrac{1}{2}\br{T^aT^bT^aT^b}$,
\item Figure \ref{fig:prop}e: $\displaystyle \br{T^aT^bT^aT^b}$,
\item Figure \ref{fig:prop}f: $\displaystyle i\dfrac{\zeta(3)}{192 \pi^3}\br{T^{a_1}T^{a_2}T^{a_3}T^{a_4}T^{a_5}T^{a_6}T^{a_6}T^{a_7}T^{a_2}T^{a_5}T^{a_1}T^{a_7}T^{a_3}T^{a_4}}$,
\end{enumerate}

The final answer in order 2 for knot $K$ is
\eq{
KI_2\br{K_{4_1}}=\dfrac{11}{12}\tr\br{T^aT^aT^bT^b}-\dfrac{11}{12}\tr\br{T^aT^bT^aT^b}
}

In orders higher than 2 the mixed terms where some chords are taken from associators and some from R-matrices start to appear. An example of such a chord configuration for order 7 is given in Figure \ref{fig:prop}f.

\section{More pictures on Vassiliev invariants}
\label{apics}
\begin{figure}[h!]
\centering\leavevmode
\includegraphics[width=15 cm]{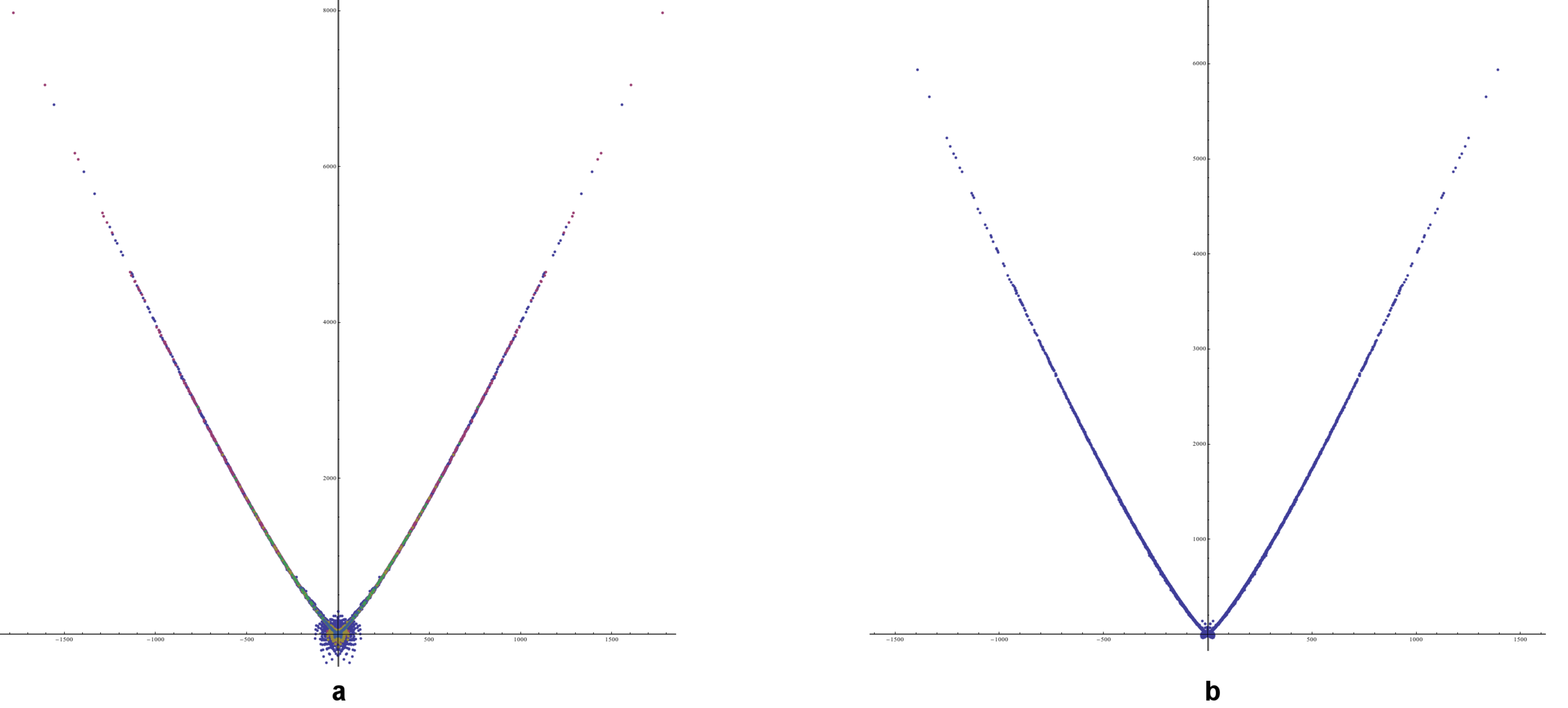}
\caption{\textbf{a}: horns, i.e. ${\cal{V}}_{6,5}$ plotted against ${\cal{V}}_{5,2}$ for knots with up to 14 crossings; \textbf{b}: same with constraint \re{constr_horns}}
\label{p_horns}
\end{figure}

Here we list several more pictures of the type of Figure \ref{p_fish}, representing ``relations'' beween Vassiliev invariants.
Very interesting is the form of the plot of ${\cal{V}}_{6,5}$ against ${\cal{V}}_{5,2}$, which is depicted in Figure \ref{p_horns}\textbf{a}. 

The following expression characterizes with good accuracy  the long ``horns'' part of the figure by vanishing on most of the corresponding knots:
\eqlm{constr_horns}{
F_3=-10752+\frac{48616 {\cal{V}}_{2,1}}{15}-178 {\cal{V}}_{2,1}^2+{\cal{V}}_{2,1}^3-1254 {\cal{V}}_{3,1}+66 {\cal{V}}_{2,1} {\cal{V}}_{3,1}-3 {\cal{V}}_{3,1}^2+356 {\cal{V}}_{4,2}-6 {\cal{V}}_{2,1} {\cal{V}}_{4,2}-\\-356 {\cal{V}}_{4,3}+6 {\cal{V}}_{2,1} {\cal{V}}_{4,3}-66 {\cal{V}}_{5,2}+66 {\cal{V}}_{5,3}+66 {\cal{V}}_{5,4}+6 {\cal{V}}_{6,5}-6 {\cal{V}}_{6,6}-6 {\cal{V}}_{6,7}-6 {\cal{V}}_{6,8},
}
see Figure \ref{p_horns}\textbf{b}, where only points on which the expression vanishes are left.

\begin{figure}[h!]
\centering\leavevmode
\includegraphics[width=15 cm]{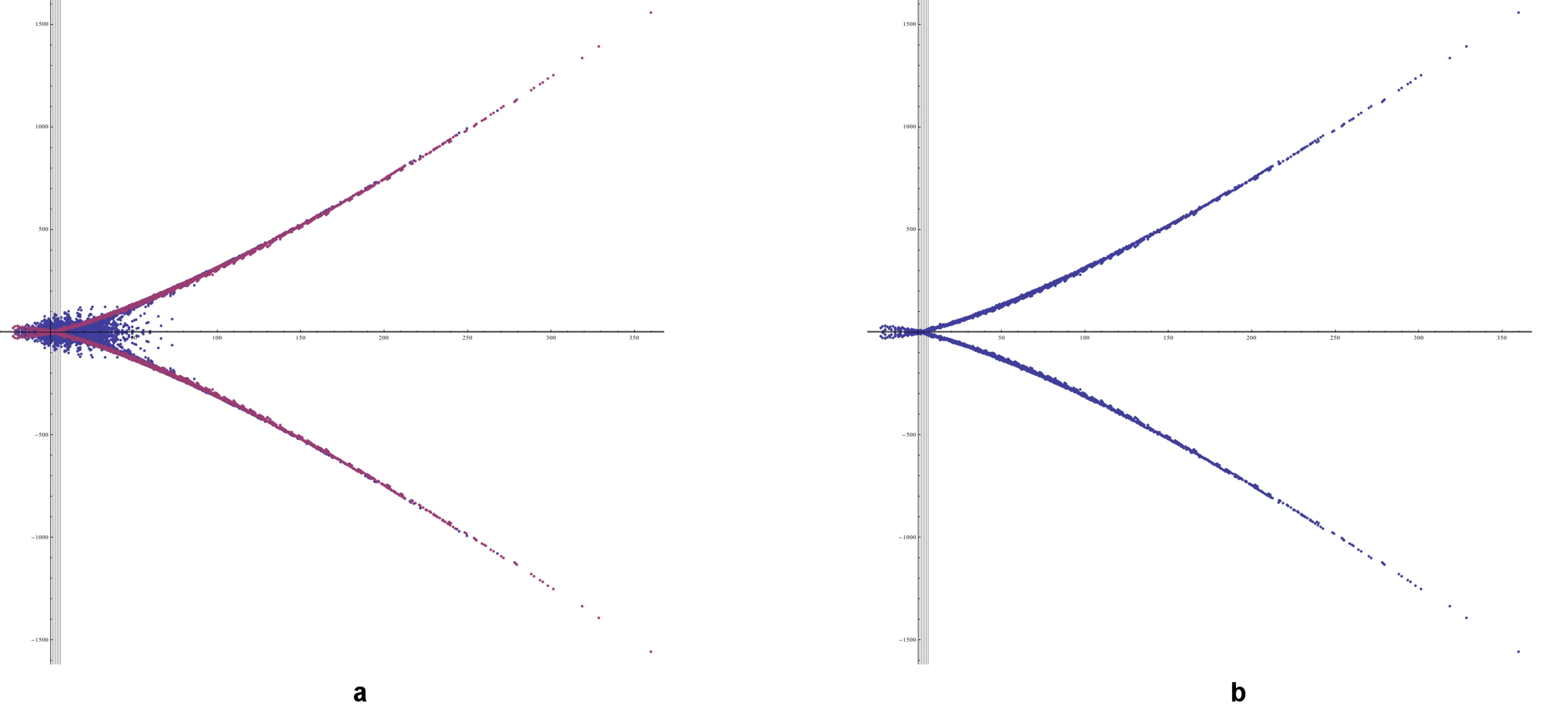}
\caption{\textbf{a}: horns, i.e. ${\cal{V}}_{5,2}$ plotted against ${\cal{V}}_{4,2}$ for knots with up to 14 crossings; \textbf{b}: same with constraint \re{constr_horns}}
\label{p_horns4}
\end{figure}

The plot of ${\cal{V}}_{5,2}$ against ${\cal{V}}_{4,2}$ has vaguely the same form, and moreover the same expression \re{constr_horns} vanishes on most of the points in the ``horns'' part, see Figures \ref{p_horns4}\textbf{a}, \ref{p_horns4}\textbf{b}.

\begin{figure}[h!]
\centering\leavevmode
\includegraphics[width=15 cm]{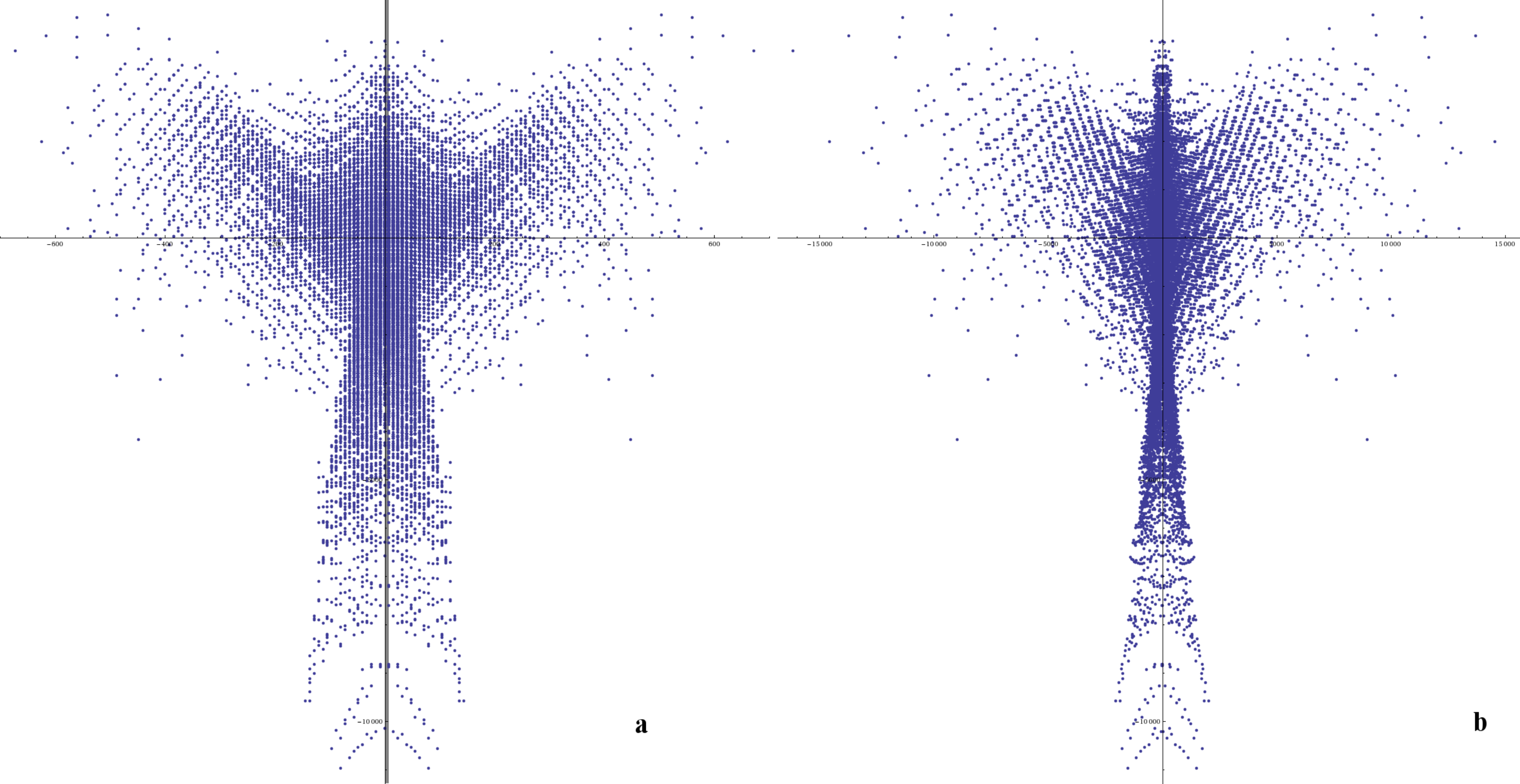}
\caption{Birds, i.e. $5 {\cal{V}}_{4,2}-31 {\cal{V}}_{4,3}$ plotted against ${\cal{V}}_{3,1}$ (Figure \textbf{a}) and $5 {\cal{V}}_{4,2}-31 {\cal{V}}_{4,3}$ plotted against ${\cal{V}}_{5,3}$ (Figure \textbf{b}), for knots with up to 14 crossings}
\label{p_horns5}
\end{figure}

Plots of $5 {\cal{V}}_{4,2}-31 {\cal{V}}_{4,3}$ against ${\cal{V}}_{3,1}$ and ${\cal{V}}_{5,3}$ are also rather interesting, see Figures \ref{p_horns5}\textbf{a}, \ref{p_horns5}\textbf{b}.

\end{document}